\documentclass[12pt]{iopart} 
\pdfoutput=1

\usepackage{supertabular}
\usepackage{indentfirst}
\usepackage{oldgerm}
\usepackage{textfit}
\usepackage{eso-pic,graphicx}
\usepackage[amssymb]{SIunits}
\usepackage{nicefrac}
\usepackage{footnote}
\usepackage{latexsym}
\usepackage{amsfonts}
\usepackage{mathrsfs}
\usepackage{bbold}
\usepackage{setstack}
\usepackage{esint}
\usepackage{nicefrac} 
\usepackage{booktabs}
\usepackage{array} 
\usepackage{paralist}
\usepackage{dsfont}
\usepackage{mathrsfs}

\usepackage{cleveref}

\newcommand{\str}{\mathrm{str}}

\newcommand{\Ber}{\mathrm{Ber}}

\eqnobysec

\begin{document} 

\title{Integrable theories and generalized graded Maillet algebras}
\author{A Melikyan$^1$ and G Weber$^2$,}
\address{$^1$ \sl{Instituto de F\'{\i}sica},
\sl{Universidade de Bras\'{\i}lia, \\70910-900, Bras\'{\i}lia, DF, Brasil}}
\address{$^1$ \sl{{International Center of Condensed Matter Physics}},\\
\sl{C.Postal 04667, Brasilia, DF, Brazil}}
\address{$^3$ \sl{Instituto de F\'{\i}sica},
\sl{Universidade de S\~{a}o Paulo,\\ C. Postal 66318, 05315-970, S\~{a}o Paulo, SP, Brasil}}
\eads{\mailto{amelik@gmail.com}, \mailto{gbrl.wbr@gmail.com}}

\begin{abstract}
We present a general formalism to investigate the integrable properties of a large class of non-ultralocal models which in principle allows the construction of the corresponding lattice versions. Our main motivation comes from the $\mathfrak{su}(1|1)$ subsector of the string theory on $AdS_5 \times S^5$ in the uniform gauge, where such type of non-ultralocality appears in the resulting  Alday-Arutyunov-Frolov ($AAF$) model. We first show how to account for the second derivative of the delta function in the Lax algebra of the $AAF$ model by modifying Maillet's $r$- and $s$-matrices formalism, and derive a well-defined algebra of transition matrices, which allows for the lattice formulation of the theory. We illustrate our formalism on the examples of the bosonic Wadati-Konno-Ichikawa-Shimizu ($WKIS$) model and the two-dimensional free massive Dirac fermion model, which can be obtained by a consistent reduction of the full $AAF$ model, and give the explicit forms of their corresponding $r$-matrices.
\end{abstract}

\pacs{02.30.Ik, 11.55.Ds, 11.25.Tq}


\maketitle

\section{Introduction}\label{sec:introduction}

The study of classical and quantum integrability of the Alday-Arutyunov-Frolov ($AAF$) model \cite{Alday:2005jm,Arutyunov:2005hd,Staudacher:2004tk,Frolov:2006cc,Arutyunov:2006ak,McLoughlin:2004dh,Swanson:2005wz,Klose:2006dd,Melikyan:2011uf,Melikyan:2012kj}, which arises in the  $\mathfrak{su}(1|1)$ sub-sector of the string theory on $AdS_5 \times S^5$ in the uniform gauge, has revealed rather non-trivial integrable properties of the model and presented new challenges to quantize it. Even though the string theory on $AdS_{5} \times S^{5}$ background is classically integrable (for a review and references, see \cite{Beisert:2010jr,Arutyunov:2009ga}), the $AAF$ model is a particularly interesting example of the fermionization technique, and serves as a representative example of  the characteristic  problems, which one will encounter trying to quantize such fermionic models. In particular, it requires a more detailed investigation and new methods to deal with  the more complex integrable structures of the non-ultralocal type in order to put the theory on a lattice. 

We have already emphasized in the previous publication \cite{Melikyan:2012kj} several, seemingly unrelated difficulties, that arise when considering the quantization of the $AAF$ model. To mention one, the attempts to investigate the quantum integrability of the $AAF$ model by utilizing the standard perturbative approach yield extremely complicated technical computations, that are not possible to carry out beyond the one-loop order considered in \cite{Melikyan:2011uf}. Moreover, the perturbative calculations do not take into account the highly non-linear Dirac bracket structure of the $AAF$ model, which makes the inverse scattering method the only reliable non-perturbative technique to quantize it.

This program has been initiated in \cite{Melikyan:2012kj}, where one of the principal results was the existence of a surprisingly simple $2 \times 2$ representation for the Lax connection. The latter made the calculation of the algebra of $L$-operators a lengthy but manageable task, which was also carried out in \cite{Melikyan:2012kj}. Moreover, it was found there that the resulting algebra is highly non-ultralocal, and contains terms proportional to the second derivative of the delta function. To our knowledge, such non-ultralocal models are quite rare, and apart from some exotic models, such as the Wadati-Konno-Ichikawa-Shimizu ($WKIS$) model, the above non-ultralocality type is new for string models. It is clear from the fermionization procedure of \cite{Alday:2005jm}, that it is due to trading the bosonic fields in favour of the fermionic ones that such non-ultralocality of higher order arises. Thus, we may expect that the similar fermionization procedure \cite{Polyakov:1983tt,Polyakov:1984et} for  other sectors of string theory will lead to the same type of higher order non-ultralocalities. It is, therefore, our main task  to give in this paper a unified formalism to deal with such cases, which allows lattice formulations of the corresponding theories as the first step towards quantization via the inverse scattering method.

The quantization of  non-ultralocal integrable systems via the standard methods has always presented difficulties, even for the much simpler models (for various methods, recent developments and applications in strings see \cite{Faddeev:1985qu,deVega:1983gy,Maillet:1985ec,Kundu:2003cu,Kundu:1996hb,Delduc:2012qb,Delduc:2012vq,Dorey:2006mx,Benichou:2010ts,Benichou:2011ch,Benichou:2012hc}). While there is no satisfactory general formalism to resolve all the difficulties of quantization, we show in this paper that by employing a suitable generalization of Maillet's $r$- and $s$-matrices formalism \cite{deVega:1983gy,Maillet:1985ec,Maillet:1985fn,Maillet:1985ek,Freidel:1991jx,Freidel:1991jv}, one can easily extract the scattering data for the $AAF$ and similar models, and obtain a well-defined algebra of monodromy matrices, which is the starting point to construct the quantum theory. In addition, one has to take into account the graded nature of the Lax pair, which makes the analysis technically more complicated. Nevertheless, as we show in this paper, the essential steps to construct a well-defined algebra of transition matrices are simple to manage, and we give the full details of this general construction. 

First, we show that the presence of higher order non-ultralocal terms amounts to a simple shift in Maillet's formalism of the $r$- and $s$-matrices \cite{Maillet:1985ek}, resulting in the corresponding $u$- and $v$-matrices, which define the algebra of transition matrices. This algebra then allows, as was shown in \cite{Freidel:1991jx,Freidel:1991jv}, a lattice formulation of such theories. Thus, the central result of this paper is that this method allows, in principle, a lattice formulation of the $AAF$ model.

We solve, as an illustrative example of our formalism, the bosonic $WKIS$ model, and give its correct $u$-matrix. We emphasize that one of the main and most non-trivial steps of this procedure is demonstrating the existence of a \emph{local} in fields $u$-matrix. The constant $u$-matrix is then obtained by considering, for example, the case of rapidly decreasing fields on the infinite line. 

In order to consider fermionic models, we also explain how to generalize all the formulas for the graded case. As a first step to deal with the full $AAF$ model, we find a rather peculiar feature of the Lax pair for the $AAF$ model. Namely, we show that it admits a consistent reduction which gives the free massive Dirac fermion theory, and has the same higher-order non-ultralocality as the full $AAF$ model. Thus, one can test our formalism on this simple and obviously integrable model. For this example, we also show the existence of a local $u$-matrix, and obtain its constant part for the case of rapidly decreasing fields. The periodic case can be treated in a similar manner, and present no difficulties due to the non-zero mass scale of the $AAF$ model \cite{Maillet:1985ek}. Moreover, we obtain the algebra of transition coefficients, which is the first step towards deriving the action-angle variables encoding the canonical structure of the model. We emphasize that the Lax pair of the free massive fermion model was obtained by a consistent reduction of the one for the full $AAF$ model, when setting the two coupling constants $g_{2}=0$ and $g_{3}=0$. Thus, the local $u$-matrix for $AAF$ model should be such that, when setting the coupling constants to zero, one obtains the formulas given in the main text. This is particularly interesting for the case of rapidly decreasing fields on the infinite line, and the corresponding constant $u$-matrix, which has a non-trivial structure already for the free fermion model. Hence, it should give some insight onto the general constant $r$-matrix of the full $AAF$ model.

Our paper is organized as follows: In section \ref{sec:ov}, we fix our notations, suitable for the later generalization for the graded case, and explain the origin of the ambiguities in the algebra of transition matrices. In section \ref{sec:gma}, we give the construction of $u$- and $v$-matrices, which generalize Maillet's formalism of the $r$- and $s$-matrices for the higher order non-ultralocal terms. We show that the algebra of transition matrices has the same functional form as in Maillet's case, with an appropriate shift of the $r$- and $s$-matrices, allowing for the formulation of the lattice version. We also explain how to construct local solutions for the $u$- and $v$-matrices, defining the algebra of the transition matrices. In section \ref{subsec:WKIS_model}, we demonstrate our formalism for the bosonic $WKIS$ model, and derive the corresponding local $u$ and $v$-matrices, as well as the constant $u$-matrix for the rapidly decreasing case. In section \ref{sec:graded_ma}, we explain how to generalize our formulas for the graded case. In section \ref{sec:aafmodel}, we briefly overview the $AAF$ model and its corresponding Lax pair, and show the existence of a consistent  reduction of it, which corresponds to the free massive Dirac fermion model. We then show the existence of a local $u(x)$-matrix, find the associated constant $u$-matrix and, as a consequence, the corresponding algebra of transition coefficients. We also explain how the local charges can be extracted from the monodromy matrix. In conclusion, we discuss the next essential steps to quantize the full $AAF$ model and comment on a possible connection between the \emph{uniform} and the \emph{uniform light-cone} gauges of the $\mathfrak{su}(1|1)$ sector of strings.  Finally, in the appendices, we give various definitions,  computational details and derivations of the principal results of the main text.

\section{Overview}\label{sec:ov}

In this section we mainly fix our notations following the standard monographs \cite{Faddeev:1987ph,Faddeev:1982rn,Novikov:1984id,Korepin:1997bk}, suitable for the generalization to the graded case, presented in section \ref{sec:graded_ma}. We start with a general two-dimensional classical field theory, defined on an interval of finite length $x \in [-L,L]$, which admits a Lax representation:
\begin{eqnarray}\label{ov:Lax_system}
	\partial_t \Psi(t,x;\lambda) &= L_0(t,x;\lambda) \Psi(t,x;\lambda),\\
	\partial_x \Psi(t,x;\lambda) &= L_1(t,x;\lambda) \Psi(t,x;\lambda). \nonumber
\end{eqnarray}
Here $\Psi(t,x;\lambda)$ is a rank $k$ vector-valued function of $t$, $x$ and the spectral parameter $\lambda$, and $L_i(t,x;\lambda)$, $i=(0,1)$ are the coefficients of the Lax connection, which depend locally on the fields and the spectral parameter $\lambda$. The equations of motion follow from the compatibility condition of the associated linear system \eref{ov:Lax_system}:
\begin{equation}\label{ov:zcc}
	\partial_t L_1(t,x;\lambda) - \partial_x L_0(t,x;\lambda) - \left[L_0(t,x;\lambda), L_1(t,x;\lambda) \right] =0.
\end{equation}

The transition matrix $T(x,y;\lambda)$ is defined as the solution of
\begin{eqnarray}
	\partial_x T(x,y;\lambda) &= L_1(x;\lambda) T(x,y;\lambda), \label{ov:diff_eq_Tx} \\ 
	\partial_y T(x,y;\lambda) &= - T(x,y;\lambda) L_1(y;\lambda), \label{ov:diff_eq_Ty}
\end{eqnarray}
with the initial condition:
\begin{eqnarray}
	\lim_{x\to y}T(x,y;\lambda) = \mathbb{1}, \label{ov:T_x_x}
\end{eqnarray}
where we omitted the time dependence and denoted by $\mathbb{1}$ the $k$-dimensional unit matrix. The transition matrix satisfies the following time evolution equation:
\begin{equation}
	\partial_t T(x,y;\lambda) = L_0(x;\lambda) T(x,y; \lambda) - T(x,y;\lambda) L_0(y,\lambda). \label{ov:T_time_evol}
\end{equation}
The monodromy matrix, defined as the transition matrix for the complete interval, $T_L(\lambda) = T(L,-L;\lambda)$, encodes all the spectral properties of an integrable model. The canonical structure of the action-angle variables can be obtained by computing the algebra of transition matrices. Namely, we need to evaluate the brackets:
\begin{eqnarray}
\left\{ T(x,y;\lambda) \stackrel{\otimes}{,} T(x',y';\mu) \right\} \label{ov:T_tens_bra}
\end{eqnarray}
for any  values of $x$, $y$, $x'$ and $y'$. Here we have employed the usual notation for the tensor product:
\begin{eqnarray}
	\left( A \otimes B \right)^{ik}_{jl} &= A^i_j B^k_l,\\
	\left\{ A \stackrel{\otimes}{,} B \right\}^{ik}_{jl} &= \left\{ A^i_j, B^k_l\right\}.
\end{eqnarray}

As is well known, the algebra \eref{ov:T_tens_bra} is not well defined when some of the points $x$, $y$, $x'$ and $y'$ coincide, if the corresponding algebra between the $L_1(z;\lambda)$-matrices, $\left\{ L_1(z;\lambda) \stackrel{\otimes}{,} L_1(z';\mu) \right\}$, is not ultralocal, i.e., contains derivatives of the delta function \cite{deVega:1983gy,Maillet:1985ec,Maillet:1985fn,Maillet:1985ek,Freidel:1991jx,Freidel:1991jv}. The presence of such non-ultralocalities leads to ambiguities in the algebra \eref{ov:T_tens_bra} which requires a regularization. The origin of the ambiguities in \eref{ov:T_tens_bra} for the non-ultralocal case can be easily seen from the following general formula \cite{Faddeev:1987ph}:
\begin{eqnarray}\label{ov:T_algebra}
	\fl \left\{ T(x,y;\lambda) \stackrel{\otimes}{,} T(x',y';\mu) \right\} &= \int\limits_y^x dz \: \int\limits_{y'}^{x'} dz'\: T(x,z;\lambda) \otimes T(x',z';\mu) \\ &\cdot \left\{ L_{1}(z;\lambda) \stackrel{\otimes}{,} L_{1}(z';\mu) \right\} \: T(z,y;\lambda) \otimes T(z',y';\mu). \nonumber
\end{eqnarray}
This equation determines the algebra of transition matrices \eref{ov:T_tens_bra} through the Lax-operator algebra, $\left\{ L_{1}(z;\lambda) \stackrel{\otimes}{,} L_{1}(z';\mu) \right\}$. In the case where some of the points coincide, for example, when $x = x'$ and $y = y'$ the formula \eref{ov:T_algebra} yields a well-defined algebra of transition matrices \eref{ov:T_tens_bra} for the ultralocal case. However, for the non-ultralocal models, i.e., the models for which the expression $\left\{ L_{1}(z;\lambda) \stackrel{\otimes}{,} L_{1}(z';\mu) \right\}$ contains derivatives of the delta function $\partial_{z}^k\delta(z-z')$, the algebra \eref{ov:T_tens_bra} following from the general formula \eref{ov:T_algebra} is not well-defined for coinciding points. For such models, the resulting algebra depends on the manner the limits $x \to x'$ and $y \to y'$ are taken. There are many models of this type and a large amount of literature devoted to solving the associated difficulties using different techniques \cite{Faddeev:1985qu,deVega:1983gy,Maillet:1985ec,Kundu:2003cu,Kundu:1996hb,Delduc:2012qb,Delduc:2012vq,Dorey:2006mx,Benichou:2010ts,Benichou:2011ch,Benichou:2012hc}.

In this paper we consider the regularization scheme based on the symmetrized Maillet bracket \cite{Maillet:1985ek}, which we discuss in the next section.

\section{Generalized Maillet Algebra}\label{sec:gma}

In this section, we consider a generalization of Maillet's formalism of $r$- and $s$-matrices to take into account non-ultralocal Lax algebras containing terms up to the second derivative of the delta function. Some partial results of this section have already been reported in the previous publication \cite{Melikyan:2012kj}. Here we give the  complete account of the derivation, and in section \ref{sec:graded_ma}, its extension to the graded case. We also refer the reader to the original work \cite{Maillet:1985ek} for a detailed exposition of the case involving the non-ultralocality only up to the first derivative of the delta function. 

Our starting point is a non-ultralocal Lax algebra of the general form:
\begin{eqnarray}\label{gma:Lax_algebra}
	\fl \left\{ L_1(z;\lambda) \stackrel{\otimes}{,} L_1(z';\mu) \right\} &= A(z;\lambda,\mu) \: \delta (z- z') + B(z;\lambda,\mu) \: \partial_z \delta (z-z') \\ 
	&+ C(z;\lambda,\mu) \: \partial_z^2 \delta(z-z'), \nonumber
\end{eqnarray}
where $A(z;\lambda,\mu)$, $B(z;\lambda,\mu)$ and $C(z;\lambda,\mu)$ are some functions of the dynamical fields. In spite of the fact that the functions $A(z;\lambda,\mu)$, $B(z;\lambda,\mu)$ and $C(z;\lambda,\mu)$ appearing in \eref{gma:Lax_algebra} depend only on the variable $z$, this is the most general Lax algebra containing terms up to the second derivative of the delta function. This is easily seen by noting that any $z'$ dependent function can be readily absorbed into $A(z;\lambda,\mu)$, $B(z;\lambda,\mu)$ and $C(z;\lambda,\mu)$ by an appropriate manipulation of the derivatives and uses of the delta functions. 
It is convenient to re-write the algebra \eref{gma:Lax_algebra} in the following form:
\begin{eqnarray}
	\left\{ L_1(z;\lambda) \stackrel{\otimes}{,} L_1(z';\mu) \right\} &= \left[ A(z;\lambda,\mu) - \frac{1}{2} \partial_z B(z;\lambda,\mu)\right] \: \delta (z-z') \label{gma:L_algebra_s1_s2} \\ 
	&- \left[ s_1(z;\lambda,\mu) + s_1(z';\lambda,\mu)\right] \: \partial_z \delta (z-z') \nonumber \\
	&+ \left[ s_2(z;\lambda,\mu) + s_2(z';\lambda,\mu)\right] \: \partial^2_z \delta (z-z') \nonumber,
\end{eqnarray}
where we have introduced the $s_{1}$- and $s_{2}$-matrices as follows:
\begin{eqnarray}
	s_1(z;\lambda,\mu) &:= -\frac{1}{2} \left[ B(z;\lambda,\mu) - \partial_z C(z;\lambda,\mu) \right], \label{gma:s1-matrix}\\
	s_2(z;\lambda,\mu) &:= \frac{1}{2} C(z;\lambda,\mu). \label{gma:s2-matrix}
\end{eqnarray}
In this case, one has to introduce an additional $s_{2}$-matrix in comparison to Maillet's original work \cite{Maillet:1985ek}. This is clearly the result of the presence of a term proportional to the second derivative of the delta function in \eref{gma:Lax_algebra}. As a matter of fact, restoring the dependence on $z'$ in an appropriate manner to pass to the notations used it \cite{Maillet:1985ek}, and setting $s_{2}=0$, we obtain Maillet's original $s$-matrix.

We now define the $r$-matrix as follows:
\begin{eqnarray}\label{gma:r-matrix}
	r(z;\lambda,\mu) := -\frac{1}{2} B(z;\lambda,\mu) + r_0(z;\lambda,\mu),
\end{eqnarray}
where $r_0(z;\lambda, \mu)$ is determined from the natural requirement that the coefficient of $\delta (z-z')$ contains the commutator term $\left[ r(z;\lambda,\mu), L_1(z;\lambda) \otimes \mathbb{1} + \mathbb{1} \otimes L_1(z;\mu)\right]$. This will allow for a manifest comparison with the ultralocal case. To this end, one first recasts the coefficient of $\delta (z-z')$ in \eref{gma:L_algebra_s1_s2} in the form:
\begin{eqnarray}\label{gma:temp_formula1}
 A(z;\lambda,\mu) - \frac{1}{2} \partial_z B(z;\lambda,\mu) =A(z;\lambda,\mu)+\partial_{z}r(z;\lambda,\mu) -\partial_{z}r_{0}(z;\lambda,\mu). 
\end{eqnarray}
As we would like the coefficient of $\delta (z-z')$ to contain the commutator term $\left[ r(z;\lambda,\mu), L_1(z;\lambda) \otimes \mathbb{1} + \mathbb{1} \otimes L_1(z;\mu)\right]$, we further set the following condition on $r_{0}(z;\lambda,\mu)$:
\begin{eqnarray}\label{gma:temp_formula2}
	A(z;\lambda,\mu)-\partial_{z}r_{0}(z;\lambda,\mu)\\
	=\left[ r(z;\lambda,\mu), L_1(z;\lambda) \otimes \mathbb{1} + \mathbb{1} \otimes L_1(z;\mu)\right] + \Lambda(z;\lambda,\mu),\nonumber
\end{eqnarray}
where the function $\Lambda(z;\lambda,\mu)$ is the extra term which accounts for the form of our choice \eref{gma:temp_formula2}, and, as we explain below, is a matrix that must depend on the fields only through $s_{1}(z;\lambda,\mu)$, $s_{2}(z;\lambda, \mu)$ and $L_1(z;\lambda)$, as well as respect the antisymmetry of the Poisson brackets. The coefficient of $\delta (z-z')$
in \eref{gma:L_algebra_s1_s2} can now be written in the form:
\begin{eqnarray}\label{gma:temp_formula1a}
 A(z;\lambda,\mu) - \frac{1}{2} \partial_z B(z;\lambda,\mu)\\ =\partial_{z}r(z;\lambda,\mu) + \left[ r(z;\lambda,\mu), L_1(z;\lambda) \otimes \mathbb{1} + \mathbb{1} \otimes L_1(z;\mu)\right] + \Lambda(z;\lambda,\mu). \nonumber
\end{eqnarray}

Using once more the equation \eref{gma:r-matrix}, one obtains from the expression above the following inhomogeneous differential equation for $r_0(z;\lambda, \mu)$:
\begin{eqnarray}\label{gma:diff_eq_r0}
	 \partial_z r_0(z;\lambda,\mu) + \left[ r_0(z;\lambda,\mu), L_1(z;\lambda) \otimes \mathbb{1} + \mathbb{1} \otimes L_1(z;\mu) \right] = \Omega(z;\lambda, \mu). 
\end{eqnarray}
The function $\Omega(z;\lambda, \mu)$ in \eref{gma:diff_eq_r0} is defined as follows:
\begin{eqnarray}\label{gma:Omega_matrix}
	 \Omega(z;\lambda, \mu) &:= A(z;\lambda, \mu) +\frac{1}{2} \left[B(z;\lambda,\mu), L_1(z;\lambda) \otimes \mathbb{1} + \mathbb{1} \otimes L_1(z;\mu) \right]\\
	& - \Lambda(z;\lambda,\mu) \nonumber.
\end{eqnarray} 
The general solution for the differential equation \eref{gma:diff_eq_r0} is:
\begin{eqnarray}\label{gma:general_r0}
	\fl r_0(x;\lambda,\mu) = \int\limits_a^x dz \: T(x,z;\lambda) \otimes T(x,z;\mu) \: \Omega(z;\lambda,\mu) \: T(z,x;\lambda) \otimes T(z,x;\mu) \\
	+  \: T(x,a;\lambda) \otimes T(x,a;\mu) \: \Omega_{0}(a;\lambda,\mu) \: T(a,x;\lambda) \otimes T(a,x;\mu), \nonumber
\end{eqnarray}
where $a \in \mathbb{R}$  and $\Omega_{0}(a;\lambda,\mu)$ is a solution of the homogeneous equation associated with \eref{gma:diff_eq_r0}, determined by the boundary conditions imposed on $r_0(x;\lambda,\mu)$. We stress that, in general, the solution for $r_0(x;\lambda,\mu)$ given by the expression \eref{gma:general_r0} is highly non-local in terms of the fields of the theory. 

Finally, substituting \eref{gma:temp_formula1a} back into the algebra \eref{gma:L_algebra_s1_s2}, it takes the following form:
\begin{eqnarray}
	 \fl \left\{ L_1(z;\lambda) \stackrel{\otimes}{,} L_1(z';\mu) \right\} &= \bigg(\partial_{z}r(z;\lambda,\mu) + \left[ r(z;\lambda,\mu), L_1(z;\lambda) \otimes \mathbb{1} + \mathbb{1} \otimes L_1(z;\mu)\right] \bigg. \label{gma:L_algebra_s1_s2_v2} \\ 
	 &\bigg.+ \Lambda(z;\lambda,\mu)\bigg) \: \delta (z-z')  \nonumber\\ 
	&- \left[ s_1(z;\lambda,\mu) + s_1(z';\lambda,\mu)\right] \: \partial_z \delta (z-z') \nonumber \\
	&+ \left[ s_2(z;\lambda,\mu) + s_2(z';\lambda,\mu)\right] \: \partial^2_z \delta (z-z') \nonumber.
\end{eqnarray}
Some comments regarding the form of this algebra are in order. First, it is convenient to make a direct comparison with the ultralocal case ($B=C=0$), in which the matrices $s_1(z;\lambda,\mu)$ and $s_2(z;\lambda,\mu)$ trivially vanish. Indeed, in this case with a constant $r$-matrix, \eref{gma:L_algebra_s1_s2_v2} reduces to the usual ultralocal algebra, provided the function $\Lambda(z;\lambda,\mu)$ vanishes. This will happen if $\Lambda(z;\lambda,\mu)$ is a function of $s_1(z;\lambda,\mu)$, $s_2(z;\lambda,\mu)$, and goes to zero as the last two functions go to zero. Secondly, one still has the freedom to choose the $\Lambda(z;\lambda,\mu)$ function, provided this restriction. The key point is that there indeed exists, as we will show below, such a $\Lambda(z;\lambda,\mu)$ function satisfying the requirement above, for which the algebra of transition matrices has a local form and reduces to the standard expression for the ultralocal case, upon setting $s_1(z;\lambda,\mu)=0$, $s_2(z;\lambda,\mu)=0$ and $\partial_z r(z;\lambda,\mu) =0$.

To find the explicit form of the $\Lambda(z;\lambda,\mu)$ function, we compute the algebra of transition matrices by substituting the Lax algebra \eref{gma:L_algebra_s1_s2_v2} into the general expression \eref{ov:T_algebra}. The explicit formulas and details of this lengthy calculation are given in \ref{app:gmadetails}. It is also shown there that the locality condition dictates the choice of the function $\Lambda(z;\lambda,\mu)$ to be of the form: 
\begin{eqnarray}\label{gma:Lambda_matrix}
	\Lambda(z;\lambda,\mu) = \left[ s_1(z;\lambda,\mu), \mathbb{1} \otimes L_1(z;\mu) - L_1(z;\lambda) \otimes \mathbb{1} \right] + h(z;\lambda,\mu),
\end{eqnarray}
where the explicit form of the function $h(z;\lambda,\mu)$ is given in \eref{gmadetails:h_functions}. Hence, the Lax algebra \eref{gma:Lax_algebra} becomes:
\begin{eqnarray}\label{gma:Lax_algebra_r_s}
	\fl \left\{ L_1(z;\lambda) \stackrel{\otimes}{,} L_1(z';\mu) \right\} &= \bigg( \partial_z r(z;\lambda,\mu) + \left[r(z;\lambda,\mu), L_1(z;\lambda) \otimes \mathbb{1} + \mathbb{1} \otimes L_1(z;\mu) \right] \bigg. \\ 
	&+ \left[ s_1(z;\lambda,\mu), \mathbb{1} \otimes L_1(z;\mu) - L_1(z;\lambda)\otimes \mathbb{1} \right] \nonumber \\
	&+ \left[ \partial_z s_2(z;\lambda,\mu), L_1(z;\lambda)\otimes\mathbb{1} + \mathbb{1} \otimes L_1(z;\mu) \right] \nonumber \\
	&+ \left[ \left[ s_2(z;\lambda,\mu), L_1(z;\lambda)\otimes \mathbb{1} \right], \mathbb{1} \otimes L_1(z;\mu) \right] \nonumber\\
	&+ \bigg. \left[ \left[ s_2(z;\lambda,\mu), \mathbb{1} \otimes L_1(z;\mu) \right], L_1(z;\lambda) \otimes \mathbb{1} \right] \bigg) \: \delta (z-z') \nonumber \\
	&- \left[ s_1(z;\lambda,\mu) + s_1(z';\lambda,\mu)\right] \: \partial_z \delta (z-z') \nonumber \\
	&+ \left[ s_2(z;\lambda,\mu) + s_2(z';\lambda,\mu)\right] \: \partial^2_z \delta (z-z') \nonumber.
\end{eqnarray}
As we show in \ref{app:gmadetails}, the algebra of transition matrices is obtained by substituting the Lax algebra \eref{gma:Lax_algebra_r_s} into the general expression \eref{ov:T_algebra}. Namely, one can show that for the points $x$, $y$, $x'$ and $y'$ all different, as discussed in the previous section, and for $x$ and $x'$ greater than $y$ and $y'$, the algebra for the transition matrices becomes:
\begin{eqnarray} \label{gma:T_algebra}
	\left\{ T(x,y;\lambda) \stackrel{\otimes}{,} T(x',y';\mu) \right\} \\
	= T(x,x_0;\lambda) \otimes T(x',x_0;\mu) \: u(x_0;\lambda,\mu) \: T(x_0,y;\lambda) \otimes T(x_0,y';\mu) \nonumber\\
	- T(x,y_0;\lambda) \otimes T(x',y_0;\mu) \: u(y_0;\lambda,\mu) \: T(y_0,y;\lambda) \otimes T(y_0,y';\mu) \nonumber\\
	+ \epsilon(x-x') \:T(x,x_0;\lambda) \otimes T(x',x_0;\mu) \: v(x_0;\lambda,\mu) \: T(x_0,y;\lambda) \otimes T(x_0,y';\mu) \nonumber\\
	+ \epsilon(y-y') \:T(x,y_0;\lambda) \otimes T(x',y_0;\mu) \: v(y_0;\lambda,\mu) \: T(y_0,y;\lambda) \otimes T(y_0,y';\mu) \nonumber,
\end{eqnarray}
where we introduced $x_0 = \min(x,x')$ and $y_0= \max(y,y')$, and defined:
\begin{eqnarray}
	\fl u(z;\lambda, \mu) := r(z;\lambda, \mu) + \partial_z s_2(z;\lambda,\mu) + \left[ s_2(z;\lambda,\mu), L_1(z;\lambda)\otimes \mathbb{1} + \mathbb{1} \otimes L_1(z;\mu)\right], \label{gma:u-matrix}\\
	\fl v(z;\lambda,\mu) := s_1(z;\lambda,\mu) + \left[ s_2(z;\lambda,\mu), L_1(z;\lambda) \otimes \mathbb{1} - \mathbb{1} \otimes L_1(z;\mu) \right]. \label{gma:v-matrix}
\end{eqnarray}


This is one of our central results, partially reported in our previous publication \cite{Melikyan:2012kj}. The algebra above has a manifestly local form, and is expressed in terms of the $u(z;\lambda, \mu)$ and $v(z;\lambda,\mu)$ matrices above. Moreover, it has exactly the same structure as Maillet's algebra (see \cite{Maillet:1985ek} for details) for the non-ultralocal terms containing only the first derivative of the delta function, if one sets $s_2(z;\lambda,\mu)=0$. Thus, Maillet's algebra, and the algebra above for higher order non-ultralocal terms have the same functional form. The latter differs from Maillet's case by a simple shift:
\begin{eqnarray}
 r(z;\lambda, \mu) \to u(z;\lambda, \mu) \quad \text{and} \quad s(z;\lambda, \mu) \to v(z;\lambda, \mu), \label{gma:shift}
 \end{eqnarray}
 where the functions $u(z;\lambda, \mu)$ and $v(z;\lambda,\mu)$ are given in \eref{gma:u-matrix} and \eref{gma:v-matrix}. Althought this suggests the existence of a gauge-equivalent Lax pair, which has the same non-ultralocality type considered by Maillet \cite{Maillet:1985ek}, it is not obvious at this point that such gauge transformation indeed exists. It would be interesting, however, to investigate this point in details.

This key observation allows one to use the formulas and techniques developed in \cite{Maillet:1985ek} to deal with the ambiguities in the algebra of transition matrices for coinciding points by simply shifting the $r(z;\lambda, \mu)$ and $s(z;\lambda, \mu)$ matrices to the matrices $u(z;\lambda, \mu)$ and $v(z;\lambda,\mu)$. In particular, one may use the same procedure to construct Maillet's symmetrized bracket, since it is based only on the form of the algebra \eref{gma:T_algebra}. One defines such a symmetrized bracket for each $n$-nested brackets as follows. First one introduces $n$-nested Poisson brackets: 
\begin{eqnarray*}
\fl \Delta^{n}(x_{i},y_{i};\lambda_{i}) := \left\{ T(x_{1},y_{1};\lambda_{1})  \stackrel{\otimes}{,} \left\{ \ldots \stackrel{\otimes}{,}\left\{ T(x_{n},y_{n};\lambda_{n})  \stackrel{\otimes}{,} \;T(x_{n+1},y_{n+1};\lambda_{n}) \right\} \ldots \right\} \right\}, 
\end{eqnarray*}
which is only a well-defined expression if all the points $x_{i}$ and $y_{i}$ are distinct. For coinciding points, one introduces the \emph{weak} Maillet brackets by a point-splitting and symmetrization regularization procedure. \footnote{The Maillet bracket is regarded \emph{weak} as its definition requires a multi-step regularization procedure. It means that a multiple bracket with a given number of factors has to be independently defined, for it is not possible to reduce its definition to a repeated regularization of multiple brackets with less factors.} For $x_{i}=x$, this means:
\begin{eqnarray}
\fl \Delta^{n}(x,y_{i};\lambda_{i}) := \lim_{\epsilon \rightarrow 0} \frac{1}{(n+1)!} \sum_{\sigma  \,  {\scriptscriptstyle\in} \, \mathbb{P}} \Delta^{n} \left(x + \epsilon \sigma(1),\ldots,x + \epsilon \sigma(n+1),y_{i};\lambda_{i} \right). \label{gma:maillet_weak_brackets}
\end{eqnarray}
Here we denoted  all possible permutations of $(1,\ldots,n+1)$ by $\mathbb{P}$. In particular, for  $n=2$, this definition leads to the Maillet bracket between the transition matrices:
\begin{eqnarray}
	\fl \{T(x,y;\lambda)  \stackrel{\otimes}{,} \;T(x,y';\mu) \}_{{M}} \nonumber\\ 
	\fl := \frac{1}{2} \lim_{\epsilon \rightarrow 0}  \left(\{T(x - \epsilon,y;\lambda)  \stackrel{\otimes}{,} \;T(x + \epsilon,y';\mu) \} + \{T(x + \epsilon,y;\lambda)  \stackrel{\otimes}{,} \;T(x - \epsilon,y';\mu) \} \right), \label{gma:M_subscript}
\end{eqnarray}
where the subscript $``M"$ denotes the Maillet bracket. 

The \emph{weak} bracket $\Delta^n(x_i,y_i;\lambda_i)$ is now a well-defined object at coinciding points and reduces to the usual Poisson bracket at non-coinciding points. More importantly, the Jacobi identity for transition matrices with coinciding points is now satisfied for this \emph{weak} bracket. Thus, allowing for a sensible computation of the algebra of monodromy matrices, which leads to the central result encoding the involution of the conserved charges:
\begin{equation}
	\left\{ \tr T_L(\lambda)^n, \tr T_L(\mu)^m \right\} = 0.
\end{equation}
Remarkably, this last bracket can be strongly defined, i.e., without the need to resort to any regularizarion. We refer the reader to the original paper \cite{Maillet:1985ek} for complete details of this construction (see also \cite{Dorey:2006mx}). We only note here that the necessity to use this construction is dictated by the presence of the $\epsilon(x-x')$ and $\epsilon(y-y')$ functions in the algebra of transition matrices \eref{gma:T_algebra}, for which there is no strong definition compatible with the Jacobi identity.

The essential property of the symmetrization procedure is that the resulting Maillet bracket \eref{gma:maillet_weak_brackets} now (weakly) satisfies the Jacobi identity. To see this, one first obtains a classical dynamical Yang-Baxter like constraint  from the Jacobi identity imposed on the original Poisson algebra of transition matrices, when all the points $x,y,x',y',x''$ and $y''$ are different. Then, one verifies that the Jacobi identity imposed on the Maillet bracket, where some of the points $x,y,x',y',x''$ and $y''$ may coincide, is satisfied, as it reduces to such classical dynamical Yang-Baxter equation. This is clearly the direct consequence of the split-point regularization procedure, utilized in the construction of Maillet bracket \eref{gma:maillet_weak_brackets}. Note, that this is true for both the original non-ultralocal algebra  considered by Maillet and the generealized algebra \eref{gma:Lax_algebra} considered in this paper. Indeed, the validity of the Jacoby identity for the Maillet bracket depends only on the form of the algebra of transition matrices \eref{gma:T_algebra}. Since both algebras have the same functional form \eref{gma:T_algebra}, differing only by the shift \eref{gma:shift}, the Jacobi identity in our case is also \emph{weakly} verified.

It is important to notice that in the original Maillet algebra the $r$- and $s$- matrices are dynamical quantities \cite{Maillet:1985ek}. The same is true in our case, the $u$- and $v$-matrices \eref{gma:u-matrix} and \eref{gma:v-matrix} are dynamical variables. In fact, given the dependence on the $r$-matrix (and consequently, on $r_0$-matrix), the $u$-matrix is usually highly non-local in terms of the fields of the theory (see equation \eref{gma:general_r0}). However, when considering, for instance, the infinite line limit, together with appropriate boundary conditions for the fields, the $u$- and $v$-matrices (just as Maillet's $r$- and $s$-matrices, in the example of the complex Sine-Gordon model considered in \cite{Maillet:1985ek}) become non-dynamical. We discuss this issue in details in section \ref{subsec:local}, and in section \ref{subsec:actang}, we consider such limit for the free fermion model explicitly obtaining non-dynamical $u_{\pm}$-matrices \eref{actang:u_plus} and \eref{actang:u_minus}, which encode the action and angle variables.

In the rest of the paper, we will use the result of \eref{gma:T_algebra} for the symmetrized algebra of transition matrices. For equal intervals $x=x'$ and $y=y'$, this result, adapted for the higher order non-ultralocal terms by performing the shift described above, reads:
\begin{eqnarray} \label{gma:T_algebra_symm_equal}
	\left\{ T(x,y;\lambda) \right. \left.\stackrel{\otimes}{,} T(x,y;\mu) \right\}_{M} &=  u(x;\lambda,\mu) \: T(x,y;\lambda) \otimes T(x,y;\mu) \\
	&- T(x,y;\lambda) \otimes T(x,y;\mu) \: u(y;\lambda,\mu).\nonumber
\end{eqnarray}
Similarly, for adjacent intervals, one finds:
\begin{eqnarray} \label{gma:T_algebra_symm_adj}
	\fl  \left\{ T(x,y;\lambda) \right. \left.\stackrel{\otimes}{,} T(y,z;\mu) \right\}_{M} =  \left( T(x,y;\lambda) \otimes \mathbb{1} \right) v(y;\lambda,\mu)\left( \mathbb{1} \otimes T(y,z;\mu) \right).
\end{eqnarray}
The formulas \eref{gma:T_algebra_symm_equal} and \eref{gma:T_algebra_symm_adj} now reveal the meaning of the $u(x;\lambda,\mu)$ and $v(x;\lambda,\mu)$ matrices. Namely, they are responsible for the equal and  adjacent intervals algebras for the transition matrices. It is interesting to note that, even though for the higher order non-ultralocality one had to introduce three independent matrices: $r(z;\lambda,\mu)$, $s_{1}(z;\lambda,\mu)$, and $s_{2}(z;\lambda,\mu)$, there are effectively only two independent matrices: $u(x;\lambda,\mu)$ and $v(x;\lambda,\mu)$, that describe the algebras \eref{gma:T_algebra_symm_equal} and \eref{gma:T_algebra_symm_adj}, and in general the algebra \eref{gma:T_algebra}. Finally, we give here the analogous formula (see for details \cite{Maillet:1985ek}) for the periodic case:
\begin{eqnarray}\label{gma:T_algebra_symm_periodic}
	 \fl \left\{ T_{L}(\lambda) \right. \left.\stackrel{\otimes}{,} T_{L}(\mu) \right\}_{M} &=  u(L;\lambda,\mu) \: T_{L}(\lambda) \otimes T_{L}(\mu) - T_{L}(\lambda) \otimes T_{L}(\mu) \: u(-L;\lambda,\mu)  \\
	&+ \left( T_{L}(\lambda) \otimes \mathbb{1} \right) v(L;\lambda,\mu) \left(\mathbb{1} \otimes  T_{L}(\mu) \right) \nonumber \\
	&- \left( \mathbb{1} \otimes T_{L}(\mu) \right) v(L;\mu,\lambda) \left(  T_{L}(\lambda)  \otimes  \mathbb{1}\right), \nonumber 
\end{eqnarray}
provided the periodicity of the fields on the interval $\left[-L,\:L\right]$. In the following sections we will use everywhere the well-defined symmetrized  Maillet brackets \eref{gma:M_subscript} omitting the explicit subscript ``M''.

\subsection{Lattice formulation: classical and quantum}

The next step towards the quantization of continuous non-ultralocal algebras of the type \eref{gma:T_algebra_symm_equal} is to formulate its lattice version following the general procedure outlined in \cite{Freidel:1991jx,Freidel:1991jv}. Clearly, the presence of derivatives of the delta function in the algebra of Lax operators still poses a problem since the Jacobi identity for transition matrices is only \emph{weakly} satisfied, whereas the consistency of the generalised quantum algebra of \cite{Freidel:1991jx,Freidel:1991jv} requires that it be \emph{strongly} satisfied. The first step in this case is to formulate a classical lattice algebra which reduces to the original one in the continuum limit.  

Indeed, the case considered here differs from \cite{Freidel:1991jx,Freidel:1991jv}, as we have shown earlier, only by the the shift \eref{gma:shift}. Therefore, one can immediately conjecture the lattice version of \eref{gma:T_algebra_symm_equal} and \eref{gma:T_algebra_symm_adj}:
\begin{eqnarray}
	\left\{ T^{n}(\lambda) \right. \left.\stackrel{\otimes}{,} T^{n}(\mu) \right\} &=  U_{1} \: T^{n}(\lambda) \otimes T^{n}(\mu) - T^{n}(\lambda) \otimes T^{n}{}(\mu) \: U_{2}, \label{gma:lat1}  \\
	\left\{ T^{n+1}(\lambda) \right. \left.\stackrel{\otimes}{,} T^{n}(\mu) \right\} &=  \left( T^{n+1}(\lambda) \otimes \mathbb{1} \right) V \:  \left( \mathbb{1} \otimes T^{n}(\mu) \right) \label{gma:lat2}\\
	\left\{ T^{n}(\lambda) \right. \left.\stackrel{\otimes}{,} T^{m}(\mu) \right\} &= 0, \quad \text{for  } |n-m| > 1. \label{gma:lat3}
\end{eqnarray}
Here, $T^{n}(\lambda) \equiv T(x_{n+1}, x_{n}; \lambda)$, and the matrices $U_{1}, U_{2}$ and $V$  may in general depend on the lattice site $n$ and on the spectral parameters $\lambda$ and $\mu$. When $V=0$, one recovers the usual ultralocal lattice version of the algebra of monodromy matrices, which in the quantum case becomes the quantum algebra:
\begin{eqnarray}
	R_{12}T^{n}_{1}T^{n}_{2} =T^{n}_{2} T^{n}_{1}R_{12}, \label{gma:RTT=TTR}
\end{eqnarray}
where the quantum $R$-matrix satisfies the usual Yang-Baxter equation.
In our case the relations \eref{gma:lat1}-\eref{gma:lat3} give rise to  more involved quantum algebras, initial study of which has been initiated in \cite{Freidel:1991jx,Freidel:1991jv}:
\begin{eqnarray}
R_{12}T^{n}_{1}T^{n}_{2} =T^{n}_{2} T^{n}_{1}\tilde{R}_{12}, \label{gma:lat1_q}\\
T^{n}_{1}T^{n+1}_{2} = T^{n+1}_{2} \hat{R}_{12}T^{n}_{1},\label{gma:lat2_q}\\
\left[ T^{n}_{1} ,T^{m}_{2} \right] = 0, \quad \text{for  } |n-m| > 1. \label{gma:lat3_q}
\end{eqnarray}

As we discussed in introduction, our principal motivation is the quantization of the $AAF$ model, which arises from  the reduction of strings on the $AdS_{5} \times S^{5}$ background to the $\mathfrak{su}(1|1)$ subsector. It was shown in \cite{Melikyan:2012kj} that the $AAF$ model admits a $2 \times 2$ Lax pair which satisfies a Lax algebra of the form \eref{gma:Lax_algebra}. Therefore, following the procedure outlined in this section, one can in principle arrive at the lattice formulation of the $AAF$ model \eref{gma:lat1}-\eref{gma:lat3} and their quantum versions \eref{gma:lat1_q}-\eref{gma:lat3_q}, as the first step towards to non-perturbative quantization. We also note here that the investigation of the algebraic structure \eref{gma:lat1_q}-\eref{gma:lat3_q} in \cite{Freidel:1991jx,Freidel:1991jv} has so far been only considered for the constant $R,\tilde{R}$ and $\hat{R}$-matrices, depending only on the spectral parameters  $\lambda$ and $\mu$. For the $AAF$ model this may not be the case, and one ends up with a more complicated algebraic structure, which may not necessarily be quadratic. This is an interesting problem that will be considered in a future publication. 

This prescription can also provide a way to put the strings on $AdS_{5} \times S^{5}$ background on a lattice by means of fermionization \cite{Polyakov:1983tt,Polyakov:1984et}. It is clear that in the general case of $AdS_{5} \times S^{5}$ string one will face the same difficulties as in the case of its $\mathfrak{su}(1|1)$ subsector, i.e., the $AAF$ model.

\subsection{Local solutions for the $u$- and $v$-matrices}\label{subsec:local}

Let us now explain how to find the \emph{local} solutions for the $u$- and $v$-matrices \eref{gma:u-matrix} and \eref{gma:v-matrix}.  We emphasize that here we are interested in local solutions for exactly  these matrices and not for the $r$-matrix, since the algebras of transition matrices \eref{gma:T_algebra}, \eref{gma:T_algebra_symm_equal} and \eref{gma:T_algebra_symm_adj} depend explicitly only on the $(u,v)$ pair.

In fact, we need to find the local solutions only for the $u$-matrix, since the $v$-matrix \eref{gma:v-matrix} is clearly local in general. This observation leads us to the following construction for the local solutions of the $u$-matrix.  Using the formulas \eref{gma:r-matrix}-\eref{gma:Lambda_matrix} one can eliminate the $r$-matrix dependence from the differential equation \eref{gma:u-matrix}, to obtain:
\begin{eqnarray}\label{local:u_diff_eq}
	\fl \partial_{z} u(z;\lambda,\mu) + \frac{1}{2}\partial_{z} B(z;\lambda,\mu) - \partial^{2}_{z} s_{2}(z;\lambda,\mu) - \partial_{z} \left[ s_{2}(z;\lambda,\mu),\:\mathcal{L}_{+}(z;\lambda,\mu) \right]  \\
	 = \Omega(z;\lambda,\mu) - \left[u(z;\lambda,\mu), \:\mathcal{L}_{+}(z;\lambda,\mu) \right] - \frac{1}{2}\left[B(z;\lambda,\mu), \:\mathcal{L}_{+}(z;\lambda,\mu) \right] \nonumber \\
	 + \left[\partial_{z}s_{2}(z;\lambda,\mu),\:\mathcal{L}_{+}(z;\lambda,\mu) \right] +  \left[\left[s_{2}(z;\lambda,\mu),\:\mathcal{L}_{+}(z;\lambda,\mu)\right],\:\mathcal{L}_{+}(z;\lambda,\mu)\right],  \nonumber
\end{eqnarray}
where we have used the notation $\mathcal{L}_{+}(z;\lambda,\mu):=L_1(z;\lambda)\otimes\mathbb{1} + \mathbb{1} \otimes L_1(z;\mu)$. 

The equation \eref{local:u_diff_eq} has in general a complex non-linear form. Each term in the right hand side of \eref{local:u_diff_eq} is, however, a local function of the fields. The  function $\Omega(z;\lambda,\mu)$, which can be computed for each concrete model from its definition \eref{gma:Omega_matrix}, can generally be represented in the form:
\begin{equation}
	\Omega(z;\lambda,\mu) = \partial_{z}\Omega_{1}(z;\lambda,\mu) + \Omega_{2}(z;\lambda,\mu), \label{local:Omega_split}
\end{equation}
where $\Omega_{1}(z;\lambda,\mu)$ and $\Omega_{2}(z;\lambda,\mu)$ are some local functions of the fields. Using \eref{local:Omega_split},  one then can trivially find a local solution to equation \eref{local:u_diff_eq} for the $u$-matrix:
\begin{eqnarray} \label{local:u_local_sol}
 u(z;\lambda,\mu) &= \Omega_{1}(z;\lambda,\mu) + \left[s_{2}(z;\lambda,\mu),\:\mathcal{L}_{+}(z;\lambda,\mu) \right] + \partial_{z}s_{2}(z;\lambda,\mu)  \\
	&-\frac{1}{2}B(z;\lambda,\mu) + \tilde{u}(\lambda,\mu), \nonumber
\end{eqnarray}
where we have denoted a coordinate \emph{independent} function as $\tilde{u}(\lambda,\mu)$, and which should satisfy, for the consistency with \eref{local:u_diff_eq}, the following equation:
\begin{equation}
	\left[\tilde{u}(\lambda,\mu),\:\mathcal{L}_{+}(z;\lambda,\mu) \right] = \Omega_{2}(z;\lambda,\mu) - \left[\Omega_{1}(z;\lambda,\mu),\:\mathcal{L}_{+}(z;\lambda,\mu)\right]. \label{local:u_tilde}
\end{equation}
Let us emphasize, that the split \eref{local:Omega_split} is not unique, and one could have absorbed some part of the first term  $\partial_{z} \Omega_{1}(z;\lambda,\mu)$ into the second term $\Omega_{2}(z;\lambda,\mu)$. Thus, there is some freedom to split the funtion $\Omega(z;\lambda,\mu)$ into $\partial_{z} \Omega_{1}(z;\lambda,\mu)$ and $\Omega_{2}(z;\lambda,\mu)$ as in \eref{local:Omega_split}. This split should, however, be done in such a way as to satisfy the consistency condition \eref{local:u_tilde} for the coordinate independent function as $\tilde{u}(\lambda,\mu)$. 

Thus, each particular model should be analysed and, as we have discussed above, the functions $\Omega_{1}(z;\lambda,\mu)$ and $\Omega_{2}(z;\lambda,\mu)$ in \eref{local:Omega_split} should be chosen in such a manner that the equation \eref{local:u_tilde} for the constant part of the $u$-matrix, $\tilde{u}(\lambda,\mu)$, has a solution. We stress that this is a non-trivial requirement, since  $\tilde{u}(\lambda,\mu)$ is a constant matrix, while the other functions appearing in \eref{local:u_tilde} are coordinate dependent. Note also that the local solution for the $u(z;\lambda,\mu)$-matrix will automatically guarantee, by means of the equation \eref{gma:u-matrix}, the locality of the corresponding $r(z;\lambda,\mu)$-matrix, even though the latter appears only in the algebra of the Lax operator \eref{gma:Lax_algebra_r_s}. Finally, for fields vanishing in the limit $z\to \pm \infty$, the algebra of transition matrices for the infinite line, i.e., the algebra \eref{gma:T_algebra_symm_adj} for equal intervals with $x=L$ and $y=-L$, in the limit $L \to \infty$ will be defined by the matrix:
\begin{eqnarray}
	\hat{u}(\lambda,\mu) = \lim_{z \to \pm \infty} u(z;\lambda,\mu). \label{local:u_hat_constant}
\end{eqnarray}
We stress that in general the matrices $\hat{u}(\lambda,\mu)$ and $\tilde{u}(\lambda,\mu)$ are different.

In the next section, we will demonstrate this method for the bosonic $WKIS$ model, and, after we explain how to deal with graded models, we will investigate, in section \ref{subsec:free}, the free fermion model, which will be obtained from a consistent reduction of the full Lax pair for the $AAF$-model. 

\section{Wadati-Konno-Ichikawa-Shimizu Model}\label{subsec:WKIS_model}

Our first illustrative example of an integrable model which displays a non-ultralocal Lax algebra of the type \eref{gma:Lax_algebra}, and which can be solved by our formalism,  is the bosonic Wadati-Konno-Ichikawa-Shimizu ($WKIS$) model \cite{wadati:1980ts,Wadati:1965ff,calogero:1982de,Tsyplyaev:1981cz,Gorder:2012rv}. There is a remarkable gauge equivalence between the generalized versions of the $WKIS$, the Hirota  and the continuous Heisenberg models \cite{Lakshmanan:1985:gs,Lakshmanan:1983sg}. Here we will only consider its simplest version, which is described by the Hamiltonian:
\begin{eqnarray}\label{wm:hamiltonian}
	H = \eta \int\limits_{-\infty}^{\infty} dx\: \left[ \sqrt{1 + \phi^*(x) \phi(x)} -1 \right],
\end{eqnarray}
where $\eta \in \mathbb{R}$ is some arbitrary constant, together with the Poisson structure:
 \begin{eqnarray}\label{wm:PBs}
	\left\{ \phi(x), \phi^*(y)\right\} = i \partial_x^2 \delta (x-y).
\end{eqnarray}
The Poisson brackets between any two functionals of the fields $F[\phi,\phi^*]$ and $G[\phi,\phi^*]$ can be trivially derived from the Poisson structure \eref{wm:PBs}, 
\begin{eqnarray}
	\left\{F,G\right\} = \iint\limits_{-\infty}^{\infty} dx \: dy\: i \partial_x^2 \delta(x-y) \:\left[ \frac{\delta F}{\delta \phi(x)} \frac{\delta G}{\delta \phi^*(y)} - \frac{\delta F}{\delta \phi^*(x)} \frac{\delta G}{\delta \phi(y)} \right].
\end{eqnarray}

Assuming $\phi(x) \in \mathscr{S}\left[ \mathbb{R} \right]$, so that $\partial_x^{n} \phi(x) \to 0$ as $|x| \to \infty$ faster than any power of $|x|^{-1}$, we obtain the Hamiltonian equation of motion for $\phi(x)$:
\begin{eqnarray}\label{wm:EoM}
	\partial_t \phi(x) = \left\{ \phi(x), H \right\} = \frac{i\eta}{2} \partial_x^2 \left[ \frac{\phi(x)}{\sqrt{1 + \phi^*(x) \phi(x)}} \right].
\end{eqnarray}
It coincides with the one studied by Wadati \cite{wadati:1980ts} for $\eta = 2$, and by Tsyplyaev \cite{Tsyplyaev:1981cz} for $\eta =1$. This equation of motion admits a Lax representation, with the following $2 \times 2$ Lax connection:
\begin{eqnarray}\label{wm:Lax_connection}
\fl L_0(z;\lambda) = \left( \begin{array}{cc}
	i\lambda^2 \frac{C}{\Gamma(z)} & 2 \lambda^2 q \sqrt{\frac{C}{\eta}} \frac{\phi(z)}{\Gamma(z)} + \lambda q \partial_z \left( \frac{\phi(z)}{\Gamma(z)} \right)\\
	-2 \lambda^2 q^* \sqrt{\frac{C}{\eta}} \frac{\phi^*(z)}{\Gamma(z)} - \lambda q^* \partial_z \left( \frac{\phi^*(z)}{\Gamma(z)} \right) & -i \lambda^2 \frac{C}{\Gamma(z)} 
	\end{array} \right), \\
	\fl L_1(z;\lambda) = -i \lambda \left( \begin{array}{cc}
		\alpha & \frac{2 q}{\eta} \phi(z) \\
		\frac{2 q^*}{\eta} \phi^*(z) & - \gamma 
		\end{array} \right),
\end{eqnarray}
where we have denoted $\Gamma(z) := \sqrt{1+\phi^*(z) \phi(z)}$, and $q \in \mathbb{C}$, $C \in \mathbb{R}^-$, $\alpha, \gamma \in \mathbb{R}$ are all constants, the last three being related by the equation:
\begin{eqnarray*}
	\alpha + \gamma = \pm 2 i \sqrt{\frac{C}{\eta}}.
\end{eqnarray*}
We note that, for the values $\alpha = \gamma = \eta = - C =1$ and $k = \frac{1}{2}$ one obtains the Lax pair given by Tsyplyaev in \cite{Tsyplyaev:1981cz}, while for the values $\alpha = \gamma =1$, $C = \eta =2$ and $k=i$ one obtains Wadati's Lax pair, given in \cite{wadati:1980ts}.

The  algebra for the $L_{1}$-operator for the Wadati model can be easily found, and has the form:
\begin{eqnarray}\label{wm:Lax_algebra}
	\left\{ L_1(z;\lambda) \stackrel{\otimes}{,} L_1(z';\mu)\right\} = -\frac{4 i |q|^2}{\eta^2} \lambda \mu \left[ \sigma_+ \otimes \sigma_- - \sigma_- \otimes \sigma_+ \right] \partial_z^2 \delta (z-z').
\end{eqnarray}
One can readily find the coefficients $A(z;\lambda,\mu)$, $B(z;\lambda,\mu)$ and $C(z;\lambda,\mu)$ by comparing the general formula \eref{gma:Lax_algebra}  with the one for the $WKIS$ model  \eref{wm:Lax_algebra}. This leads to the following $r$- and $s$-matrices for the $WKIS$ model:
\begin{eqnarray}
	r(z;\lambda,\mu) &= r_0(z;\lambda,\mu),\\
	s_1(z;\lambda,\mu) &= 0,\\
	s_2(z;\lambda,\mu) &=  -\frac{2 i |q|^2}{\eta^2} \lambda \mu \left[ \sigma_+ \otimes \sigma_- - \sigma_- \otimes \sigma_+ \right].
\end{eqnarray}

As a consequence, the expression \eref{gma:Omega_matrix} for the function $\Omega(z;\lambda, \mu)$ simplifies, and takes the form:
\begin{eqnarray}\label{wm:Omega_matrix_WKIS}
	\Omega(z;\lambda, \mu) = &- \left[ \left[ s_2(z;\lambda, \mu), L_1(z;\lambda) \otimes \mathbb{1} \right], \mathbb{1} \otimes L_1(z;\mu) \right] \\ 
	&- \left[ \left[ s_2(z;\lambda, \mu), \mathbb{1} \otimes L_1(z;\mu) \right] ,L_1(z;\lambda) \otimes \mathbb{1} \right]. \nonumber
\end{eqnarray}
Using the results of section \ref{subsec:local}, where we have explained how to construct the local solutions for the $u(z;\lambda, \mu)$-matrix, one finds from \eref{local:u_local_sol} and \eref{local:u_tilde} the equations determining the local $u(z;\lambda, \mu)$-matrix of the $WKIS$ model:
\begin{eqnarray}
		u(z;\lambda,\mu) = \left[s_{2}(z;\lambda,\mu),\:\mathcal{L}_{+}(z;\lambda,\mu) \right] +  \tilde{u}(\lambda,\mu),\label{wm:u_local_sol} \\
		\left[\tilde{u}(\lambda,\mu),\:\mathcal{L}_{+}(z;\lambda,\mu) \right] = \Omega_{2}(z;\lambda,\mu),\label{wm:utilde}
\end{eqnarray}
where we have chosen $\Omega_{1}(z;\lambda, \mu)=0$ and, therefore, $\Omega_{2}(z;\lambda, \mu)=\Omega(z;\lambda, \mu)$, given in \eref{wm:Omega_matrix_WKIS}.

Remarkably, one can show that the solution of the equation \eref{wm:utilde} for the constant part $\tilde{u}(\lambda,\mu)$ indeed exists, and has the form:
\begin{eqnarray}\label{wm:u_tilde-matrix}
	\tilde{u}(\lambda,\mu) = \tilde{r}\: \mathbb{1} \otimes \mathbb{1} &+ \frac{2 |q|^2}{\eta^2} \lambda \mu \left\{ -\frac{(\alpha + \gamma)\lambda\mu}{\lambda-\mu} (\mathbb{1} \otimes \mathbb{1} - \sigma_3 \otimes \sigma_3) \right.\\ 
	&+ \left. (\alpha + \gamma) \frac{\lambda^2 + \mu^2}{\lambda -\mu} (\sigma_+ \otimes \sigma_- + \sigma_-\otimes \sigma_+) \right\}.\nonumber
\end{eqnarray}
Then, using the equation \eref{wm:u_local_sol}, one finally finds the following local solution for the $u(z;\lambda,\mu)$-matrix:

\begin{eqnarray}\label{wm:u-matrix}
	u(x;\lambda,\mu) = \tilde{r}\: \mathbb{1} \otimes \mathbb{1} &+ \frac{2 |q|^2}{\eta^2} \lambda \mu \left\{ -\frac{(\alpha + \gamma)\lambda\mu}{\lambda-\mu} (\mathbb{1} \otimes \mathbb{1} - \sigma_3 \otimes \sigma_3) \right.\\ 
	&+ \left. (\alpha + \gamma) \frac{\lambda^2 + \mu^2}{\lambda -\mu} (\sigma_+ \otimes \sigma_- + \sigma_-\otimes \sigma_+) \right. \nonumber \\
	&+ \left. \frac{2}{\eta} \left[ -\lambda q^* \phi^*(x) \sigma_3 \otimes \sigma_- + \mu q^* \phi^*(x) \sigma_- \otimes \sigma_3 \right. \right. \nonumber \\
	&- \left. \left. \lambda q \phi(x) \sigma_3 \otimes \sigma_+ +\mu q \phi(x) \sigma_+ \otimes \sigma_3 \right] \right\}, \nonumber
\end{eqnarray}
where $\tilde{r}$ is some arbitrary constant depending only on the spectral parameter, which can be set to zero without any loss of generality. As we have remarked before, the $v(z;\lambda,\mu)$-matrix is always a local function, and can be easily found from the equation \eref{gma:v-matrix}:
\begin{eqnarray}  \label{wm:v-matrix}
	v(x; \lambda,\mu) &=  \frac{2 |q|^2}{\eta^2} \lambda \mu \left\{ (\alpha + \gamma) (\lambda + \mu) (\sigma_+ \otimes \sigma_- + \sigma_-\otimes \sigma_+) \right. \\
	&- \left. \frac{2}{\eta} \left[ \lambda q^* \phi^*(x) \sigma_3 \otimes \sigma_- + \mu q^* \phi^*(x) \sigma_- \otimes \sigma_3 \right. \right. \nonumber \\
	&+ \left. \left. \lambda q \phi(x) \sigma_3 \otimes \sigma_+ +\mu q \phi(x) \sigma_+ \otimes \sigma_3 \right] \right\}. \nonumber
\end{eqnarray}

Finally, we comment that using these results, one can consider, for example, the infinite line limit, with the fields $\phi(z) \to 0$ when $z\to \pm \infty$. In this case, the constant $\tilde{u}(\lambda,\mu)$-matrix will encode the action-angle variables, which can be constructed without any difficulties, following the standard procedures described in the monographs \cite{Faddeev:1987ph,Faddeev:1982rn,Novikov:1984id,Korepin:1997bk}. The action-angle variables for the WKIS model were also studied in the earlier paper \cite{Tsyplyaev:1981cz}. However, our results differ by some constants, which we believe is the result of not taking into account in \cite{Tsyplyaev:1981cz} the subtleties related to the ambiguity of the transition matrices.

\section{Graded Maillet Algebra}\label{sec:graded_ma}

In this section we discuss the generalization of the results obtained in section \ref{sec:gma} to the case of graded Lax pairs, which is necessary to investigate the $AAF$ model. In particular, the main goal is to carefully deal with graded Lax operators, and their tensor products, which will appear when considering the Poisson brackets of transition matrices and when analysing the Jacobi identity. 

The main motivation of our paper, the $AAF$ model, arises from  the reduction of strings on the $AdS_{5} \times S^{5}$ background to the $\mathfrak{su}(1|1)$ subsector. Let us recall that the general Lax pair for strings on the $AdS_{5} \times S^{5}$ background has the following graded structure \cite{Beisert:2010jr,Arutyunov:2009ga,Alday:2005jm,Arutyunov:2005hd,Alday:2005gi}:\footnote{Let us remind that the superalgebra $\mathfrak{psu}(2,2|4)$, defined as the quotient algebra of $\mathfrak{su}(2,2|4)$ over a $\mathfrak{u}(1)$ factor, has no realization in terms of $8\times 8$ matrices. So, strictly speaking $\mathscr{L}_{\alpha} \in \mathfrak{su}(2,2|4)$. However, we write it as an element of $\mathfrak{psu}(2,2|4)$ in order to remind that any two elements (matrices) that differ by multiples of the identity matrix should be identified. This is especially crucial when one considers the zero-curvature condition \cite{Alday:2005gi,Melikyan:2012kj}.}
\begin{eqnarray}
	\mathscr{L}_{\alpha} &= \left( \begin{array}{cccc} 
		A & F  \\
	 G & B
		\end{array} \right). \label{graded_ma:Lax_strings}
\end{eqnarray}
Here the $(4 \times 4)$ matrices $A$ and $B$ are even, and the $(4 \times 4)$ matrices $F$ and $G$ are odd.

The Lax pair for the $AAF$ model inherits this graded structure. For example, the $L_{1}$ operator has the following explicit form (see \ref{app:Laxpair} for details, and explicit dependence on the fields):
\begin{eqnarray}
	L_{1}&= \left( \begin{array}{cccc} 
		\xi_{0}^{\scriptscriptstyle{(\sigma)}} + \xi_{1}^{\scriptscriptstyle{(\sigma)}} & \Lambda^{\scriptscriptstyle{(-)}}_{\sigma}  \\
	 \Lambda^{\scriptscriptstyle{(+)}}_{\sigma} & \xi_{0}^{\scriptscriptstyle{(\sigma)}} - \xi_{1}^{\scriptscriptstyle{(\sigma)}}
		\end{array} \right), \label{graded_ma:Lax_L1_aaf}
\end{eqnarray}
where the diagonal elements $\xi_{0}^{\scriptscriptstyle{(\sigma)}}$ and $\xi_{1}^{\scriptscriptstyle{(\sigma)}}$ are even, and the anti-diagonal elements $\Lambda^{\scriptscriptstyle{(-)}}_{\sigma}$ and $\Lambda^{\scriptscriptstyle{(+)}}_{\sigma}$ are odd.

When considering the algebra of transition matrices for the graded case, one has to generalize the construction from the bosonic case, which requires the concept of the \emph {supertensor product}. It has similar properties to the usual tensor product, such as
\begin{equation}
	(A \otimes_s B)(C \otimes_s D) = AC \otimes_s BD,
\end{equation}
allowing one to write various formulas in a compact and closed form, in a manner analogous to the bosonic case. For the detailed mathematical construction and some relevant applications in the context of integrable models, we refer the reader to the monograph \cite{Berezin:1987ue} and the original papers \cite{Kulish:1980ii,Kulish:1985bj,Gohmann:1998sm,Gohmann:1999av,Gohman:2002kp} (for a comprehensive review, see \cite{Essler:2005bk}). Here we only state the main result, which is the direct generalization of the formula \eref{ov:T_algebra} for the algebra of transition matrices to the graded case:
\begin{eqnarray}\label{graded_ma:T_algebra}
	\fl \left\{ T(x,y;\lambda) \stackrel{\otimes_s}{,} T(x',y';\mu) \right\} &= \int\limits_y^x dz \: \int\limits_{y'}^{x'} dz'\: T(x,z;\lambda) \otimes_s T(x',z';\mu) \\ &\cdot \left\{ L(z;\lambda) \stackrel{\otimes_s}{,} L(z';\mu) \right\} \: T(z,y;\lambda) \otimes_s T(z',y';\mu). \nonumber
\end{eqnarray}
As in the bosonic case, this formula is valid only for $x$, $y$, $x'$ and $y'$ all different.  

Thus, we come to the following conclusion: all the formulas from sections \ref{sec:ov} and \ref{sec:gma} can be generalized to the graded case without any essential modification, by simply replacing everywhere the usual tensor product by the supertensor product. We will not write down the same expressions here, and simply use the appropriate formulas from sections \ref{sec:ov} and \ref{sec:gma} by making this change.  
We will use these formulas, when considering the fermionic $AAF$ model and its reduction to the free fermion model in the subsequent sections, as well as in the analysis of the algebra of transition coefficients and the local charges.

\section{Alday-Arutyunov-Frolov model: $\mathfrak{su}(1|1)$ strings in uniform gauge}\label{sec:aafmodel}

The Alday-Arutyunov-Frolov ($AAF$) model \cite{Alday:2005jm} arises in the $\mathfrak{su}(1|1)$ sector of the superstring theory on $AdS_{5} \times S^{5}$ in the \emph{uniform gauge}, and is one of the most non-trivial integrable fermionic models. For its derivation and various discussions we refer to the works \cite{Arutyunov:2005hd,Staudacher:2004tk,Frolov:2006cc,Arutyunov:2006ak,McLoughlin:2004dh,Swanson:2005wz,Melikyan:2011uf,Melikyan:2012kj}. The $AAF$ model provides an example of the fermionization technique, and has several interesting characteristic features. First, it is a purely fermionic relativistic two-dimensional model with a complicated interaction structure, containing terms up to the sixth order in the fermionic fields, and which is classically integrable. The perturbation theory was studied in \cite{Klose:2006dd} and in \cite{Melikyan:2011uf}. The S-matrix factorization was shown up to the one-loop order in \cite{Melikyan:2011uf}, thus, indicating the quantum integrability of the model. The full analysis within the perturbation theory, however, is beyond reach, due to formidable computational difficulties. Secondly, the attempts to apply the standard methods of integrable models fail as well, due to the presence of non-ultralocal terms up to the second order derivative in the delta function. Hence, the $AAF$ model provides an interesting example to apply our formalism. 

This study was initiated in the previous publication \cite{Melikyan:2012kj}, where it was shown that the original $4 \times 4$ Lax representation \cite{Alday:2005jm} can be surprisingly reduced to a $2 \times 2$ Lax connection. This result made the computation of the functions $A(z;\lambda,\mu)$, $B(z;\lambda,\mu)$ and $C(z;\lambda,\mu)$  in the algebra of Lax operators \eref{gma:Lax_algebra} a feasible task. In fact, the explicit form of all these functions has already been found in the previous publication \cite{Melikyan:2012kj}. Moreover, one can easily compute the function $\Omega(z;\lambda,\mu)$ from \eref{gma:Omega_matrix}, which makes it possible to apply the procedure of finding the local solutions for the $u$-matrix, as we have explained in the section \ref{subsec:local}. It is, however, very instructive to apply first our formalism to the model obtained by a consistent reduction of the full Lax pair for the $AAF$ model. This model turns out to be the free fermion model, which has all the characteristic features to make our formalism, discussed in sections \ref{sec:gma} and \ref{sec:graded_ma}, useful, and, on the other hand, makes the calculations less tedious as will be the case for the full $AAF$ model, which will be done in a subsequent publication. 

Besides the calculational simplification the reduced model provides the limiting case of the full $AAF$ model. Thus, after the results for the full $AAF$ model are obtained, one should be able to recover, at each step of the procedure, the corresponding results for the reduced model, which we give in full details in this paper. The most important open question is of course the relation of the $(u,v)$-pair of the full $AAF$ model to the one of the reduced model - the free fermion model. There is an interesting connection between the Lax pair of the reduced model, and the one obtained in $\mathfrak{su}(1|1)$ sector of the strings in the \emph{uniform light-cone gauge} \cite{Arutyunov:2005hd}. The latter turns out to be a free fermion model as well, and, in fact, the two Lax pairs (one obtained in this paper from the reduction of the full $AAF$ model, and the other obtained from the $AdS_{5} \times S^{5}$ Lax pair in the uniform light-cone gauge) coincide. Thus, it will be very interesting to compare the $(u,v)$-pair of the full $AAF$ model to the one of the reduced model. 

We give here only the main formulas and definitions, and refer the reader for complete technical details to the works \cite{Alday:2005jm,Melikyan:2011uf,Melikyan:2012kj}. The Lagrangian for the $AAF$ model has the form, see \ref{app:Laxpair} for notations:
\begin{eqnarray}
	\fl \label{aafmodel:Lagrangian} \mathscr{L} =  -J-\frac{i}{2} \left(\bar{\psi} \rho^0 \partial_0 \psi -\partial_0 \bar{\psi} \rho^0 \psi \right) + i \frac{k}{2J} \left(\bar{\psi}\rho^1 \partial_1 \psi -\partial_1 \bar{\psi} \rho^1 \psi \right) + \bar{\psi}\psi  \\
 +\frac{k \: g_{2}}{4 J^{4}} \epsilon^{\alpha \beta} \left( \bar{\psi}
\partial_{\alpha} \psi \; \bar{\psi} \rho^5 
\partial_{\beta} \psi-\partial_{\alpha}\bar{\psi} \psi \; 
\partial_{\beta} \bar{\psi} \rho^5 \psi \right) - \frac{k \: g_{3}}{16 J^{3}} \epsilon^{\alpha \beta} \left(\bar{\psi}\psi\right)^2 \partial_{\alpha}\bar{\psi}\rho^5 \partial_{\beta}\psi. \nonumber
\end{eqnarray}
Here, $k=\sqrt{\lambda}$, where $\lambda$ is the t'Hooft coupling, and $J$, the total angular momentum of the string in $S^5$. The coupling constants $g_{2}$ and $g_{3}$ should satisfy the condition $g_{2}^{2} = g_{3}$ (for its derivation, see \cite{Melikyan:2011uf}), which not only guarantees the S-matrix factorization up to the one-loop order, but also the classical integrability of the model, as was shown in \cite{Melikyan:2012kj}. One can remove both coupling constants $g_{2}$ and $g_{3}$ in the classical theory by an appropriate rescaling of the fermionic fields. We will assume everywhere below such rescaling and set the coupling constants $g_{2}=1$ and $g_{3}=1$. 

We will use here the $2 \times 2$ Lax connection obtained in \cite{Melikyan:2012kj}. We only comment that it was derived through the analysis of the equations following from the zero-curvature condition applied to the $4 \times 4$ Lax representation of \cite{Alday:2005jm}, and by noting that each equation appeared exactly twice. This allowed us to reorganize the elements of the Lax connection so that the equations became independent, reducing the Lax connection to a simpler $2 \times 2$ representation. It has the following form:
\begin{eqnarray}
	\fl	L_{0}(x;\lambda) &= \xi_{0}^{\scriptscriptstyle{(\tau)}}(x;\lambda) \mathbb{1} + \xi_{1}^{\scriptscriptstyle{(\tau)}}(x;\lambda) \sigma^{3} + \Lambda^{\scriptscriptstyle{(-)}}_{\tau}(x;\lambda) \sigma^+ + \Lambda^{\scriptscriptstyle{(+)}}_{\tau}(x;\lambda) \sigma^-, \label{aafmodel:Lax_pair_L0} \\
	\fl L_{1}(x;\lambda) &= \xi_{0}^{\scriptscriptstyle{(\sigma)}}(x;\lambda) \mathbb{1} + \xi_{1}^{\scriptscriptstyle{(\sigma)}}(x;\lambda) \sigma^{3} + \Lambda^{\scriptscriptstyle{(-)}}_{\sigma}(x;\lambda) \sigma^+ + \Lambda^{\scriptscriptstyle{(+)}}_{\sigma}(x;\lambda) \sigma^-, \label{aafmodel:Lax_pair_L1}
\end{eqnarray}
where $\sigma^i$, $i=+,-,3$ are the  Pauli matrices. The explicit form of the functions $\xi_{j}^{\scriptscriptstyle{(\sigma,\tau)}}(x;\lambda)$, $j=0,1$ and $\Lambda_{\sigma,\tau}^{\scriptscriptstyle{(\pm)}}(x;\lambda)$ is given in \ref{app:Laxpair}. The functions $\xi_{j}^{\scriptscriptstyle{(\sigma,\tau)}}(x;\lambda)$ are bosonic, and the functions $\Lambda_{\sigma,\tau}^{\scriptscriptstyle{(\pm)}}(x;\lambda)$, fermionic. Thus, the Lax connection $L_{\alpha}(x;\lambda)$ is an even supermatrix. 

The equations of motion for the $AAF$ model follow from the Lax pair $L_{0}(x;\lambda)$ and $L_{1}(x;\lambda)$ in \eref{aafmodel:Lax_pair_L0} and \eref{aafmodel:Lax_pair_L1}, and are obtained from the anti-diagonal elements of the zero-curvature condition, while the diagonal elements produce highly non-trivial identities \cite{Melikyan:2012kj}. Moreover, the Dirac brackets were computed in \cite{Alday:2005jm,Melikyan:2012kj}, and shown to have an ultralocal form. However, as was shown in \cite{Melikyan:2012kj}, the algebra for the Lax connection  $L_{1}(x;\lambda)$ in  \eref{aafmodel:Lax_pair_L1} has precisely the form \eref{gma:Lax_algebra}, discussed in sections \ref{sec:gma} and \ref{sec:graded_ma}. Thus, our formalism can be applied, and, in particular, it should be straightforward to find the $u$- and $v$-matrices, defining the algebra of the transition matrices for the infinite line and periodic cases, in a manner similar to the case of the $WKIS$ model (see section \ref{subsec:WKIS_model}).


In the case of the $AAF$ model additional technical complications arise due to the extremely involved structure of the Lax connection (see \ref{app:Laxpair} for the explicit form). In addition, one needs to take into account the graded structure of the Lax connection, and use the appropriate formulas as explained in section \ref{sec:graded_ma}. We postpone this analysis for a future publication and consider instead a simpler model, obtained by a consistent reduction of the full $AAF$ Lax connection \eref{aafmodel:Lax_pair_L0} and \eref{aafmodel:Lax_pair_L1}, which allows one to fully illustrate the formalism of sections \ref{sec:gma} and \ref{sec:graded_ma} in details, without the tedious computational complications of the $AAF$ model. It turns out, that this reduced model is the free massive Dirac fermion theory. In the sections below we describe its rather non-trivial Lax connection, and derive the algebra of transition coefficients. We stress that the similar calculation for the $AAF$ model should only result in more lengthy and tedious computations, but otherwise will not have any principal differences from the free fermion case.

\subsection{Reduction to free fermions}\label{subsec:free}

As we mentioned in the previous section, the Lagrangian \eref{aafmodel:Lagrangian} contains two coupling constants $g_{2}$ and $g_{3}$, which should satisfy the non-trivial relation $g_{2}^{2}=g_{3}$ in order to guarantee the S-matrix factorization and classical integrability. It is interesting, however, to investigate the simplest case by setting $g_{2}=0$ and $g_{3}=0$. Thus, the $AAF$ Lagrangian reduces to that of the massive free fermionic Dirac model:
\begin{eqnarray}
 \mathscr{L} &=  -J-\frac{i}{2} \left(\bar{\psi} \rho^0 \partial_0 \psi -\partial_0 \bar{\psi} \rho^0 \psi \right) + i \frac{k}{2J} \left(\bar{\psi}\rho^1 \partial_1 \psi -\partial_1 \bar{\psi} \rho^1 \psi \right) + \bar{\psi}\psi. \label{free:Lagrangian}
\end{eqnarray}
The equations of motions for free fermions, in terms of the components $\chi_{i}$ (see \eref{notations:chis} in \ref{app:Laxpair}), have the form:
\begin{eqnarray}
	\dot{\chi}_{1} &= i \chi_{1} - \frac{ik}{J}\chi_{2}',\quad  \dot{\chi}_{2} = -i \chi_{2} + \frac{ik}{J}\chi_{1}', \label{free:eoms12} \\
	\dot{\chi}_{3} &= -i \chi_{3} +\frac{ik}{J}\chi_{4}', \quad \dot{\chi}_{4} = i \chi_{4} - \frac{ik}{J}\chi_{3}' .\label{free:eoms34}
\end{eqnarray}
Note that one can rescale the world-sheet coordinate $\sigma$: $\sigma \to -\frac{k}{J}\sigma$, to write the corresponding action in the explicitly relativistic two-dimensional invariant form \cite{Alday:2005jm,Klose:2006dd,Melikyan:2011uf}.

This reduction to the free fermion case can also be seen from the full Lax pair \eref{aafmodel:Lax_pair_L0} and \eref{aafmodel:Lax_pair_L1}. It is easy to understand how to make such consistent reduction. We have noted above that the equations of motion for the $AAF$ model follow from the anti-diagonal elements of the zero-curvature condition,\footnote{To emphasize the technical complications of the full $AAF$ model, we note that the corresponding equations of motions are very involved, containing terms up to the seventh order in the fields and their space derivatives.} while the diagonal elements give several non-trivial constraints, satisfied on the equations of motion \cite{Melikyan:2012kj}. Since in the free fermion case \eref{free:Lagrangian} the equations of motion \eref{free:eoms12} and \eref{free:eoms34} are linear, one should restrict the anti-diagonal elements of the Lax connection to terms linear in the fields, while for the diagonal ones only terms up to second order should be kept. This can be explicitly verified by the direct calculation. Namely, using the formulas from \ref{app:Laxpair}, the corresponding reduced Lax pair for the free fermion takes the simpler form:
\begin{eqnarray}
		 \mathcal{L}_{0}(x;\lambda) &= \hat{\xi}_{0}^{\scriptscriptstyle{(\tau)}}(x;\lambda) \mathbb{1} + \hat{\xi}_{1}^{\scriptscriptstyle{(\tau)}}(x;\lambda) \sigma^{3} + \hat{\Lambda}^{\scriptscriptstyle{(-)}}_{\tau}(x;\lambda) \sigma^+ + \hat{\Lambda}^{\scriptscriptstyle{(+)}}_{\tau}(x;\lambda) \sigma^-, \label{free:Lax_pair_L0} \\
	 \mathcal{L}_{1}(x;\lambda) &= \hat{\xi}_{0}^{\scriptscriptstyle{(\sigma)}}(x;\lambda) \mathbb{1} + \hat{\xi}_{1}^{\scriptscriptstyle{(\sigma)}}(x;\lambda) \sigma^{3} + \hat{\Lambda}^{\scriptscriptstyle{(-)}}_{\sigma}(x;\lambda) \sigma^+ + \hat{\Lambda}^{\scriptscriptstyle{(+)}}_{\sigma}(x;\lambda) \sigma^- \label{free:Lax_pair_L1}.
\end{eqnarray}
The functions $\hat{\xi}_{j}^{\scriptscriptstyle{(\sigma,\tau)}}(x;\lambda)$, $j=0,1$, and $\hat{\Lambda}_{\sigma,\tau}^{\scriptscriptstyle{(\pm)}}(x;\lambda)$ have the following explicit form:
\begin{eqnarray}
	\hat{\xi}^{\scriptscriptstyle{(\sigma)}}_0 &= \frac{1}{4J} \left[ - \chi_3\chi_1' + \chi_4\chi_2' - \chi_1\chi_3' + \chi_2\chi_4' \right], \label{free:xis0}  \\ 
	\hat{\xi}^{\scriptscriptstyle{(\sigma)}}_1 &= \frac{il_2 J}{2 k}, \label{free:xis1} \\
\hat{\Lambda}^{\scriptscriptstyle{(-)}}_{\sigma} &= \frac{1}{\sqrt{J}} \left[ -l_{3}\chi_2'- il_{4}\chi_1' \right], \label{free:Lambdasm}  \\	
	\hat{\Lambda}^{\scriptscriptstyle{(+)}}_{\sigma} &= \frac{1}{\sqrt{J}} \left[ -l_{3}\chi_4'+ il_{4}\chi_3' \right],\label{free:Lambdasp}
\end{eqnarray}
and:
\begin{eqnarray}
	\hat{\xi}^{\scriptscriptstyle{(\tau)}}_0 &= \frac{i}{2J} \left[\chi_{3}\chi_{1}+\chi_{4}\chi_{2} \right]+ \frac{1}{4J}\left[-\chi_{3}\dot{\chi}_{1}-\chi_{1}\dot{\chi}_{3} +\chi_{4}\dot{\chi}_{2}  +\chi_{2}\dot{\chi}_{4} \right],  \\
	\hat{\xi}^{\scriptscriptstyle{(\tau)}}_1 &= -\frac{i l_{1}}{2}, \\
	\hat{\Lambda}^{\scriptscriptstyle{(-)}}_{\tau} &= -\frac{i}{\sqrt{J}}\left[l_{3}\chi_{2} - il_{4}\chi_{1} \right] - \frac{1}{\sqrt{J}} \left[l_{3}\dot{\chi}_{2} + il_{4}\dot{\chi}_{1}  \right],  \\
	\hat{\Lambda}^{\scriptscriptstyle{(+)}}_{\tau} &= \frac{i}{\sqrt{J}}\left[l_{3}\chi_{4} + il_{4}\chi_{3} \right] - \frac{1}{\sqrt{J}} \left[l_{3} \dot{\chi}_{4} - il_{4}\dot{\chi}_{3}  \right].
\end{eqnarray}

The Lax algebra \eref{gma:Lax_algebra} can be easily found, and has the form:
\begin{eqnarray}\label{free:Lax_algebra}
	\fl \left\{ \mathcal{L}_{1}(z;\lambda) \stackrel{\otimes_{s}}{,} \mathcal{L}_{1}(z';\mu) \right\} &= A(z;\lambda,\mu) \: \delta (z- z') + B(z;\lambda,\mu) \: \partial_z \delta (z-z') \\ 
	&+ C(z,;\lambda,\mu) \: \partial_z^2 \delta(z-z'), \nonumber
\end{eqnarray}
where the functions $A(z;\lambda,\mu)$, $B(z;\lambda,\mu)$ and $C(z;\lambda,\mu)$ are given by:
\begin{eqnarray}
	\fl A(z;\lambda,\mu) = \frac{i}{8J^{2}}\left[\chi_{1}\chi_{3}'' + \chi_{3}\chi_{1}'' + \chi_{4}\chi_{2}'' +\chi_{2}\chi_{4}'' \right]\left(\mathbb{1} \otimes \mathbb{1} \right) \label{free:A} \\ 
	\fl + \frac{1}{2 J^{\scriptscriptstyle{3/2}}}\left[-i l_{3}(\lambda) \chi_{2}'' - l_{4}(\lambda)\chi_{1}''\right]\left(\sigma^{+} \otimes \sigma^{3} \right)
	+ \frac{1}{2 J^{\scriptscriptstyle{3/2}}}\left[-i l_{3}(\lambda) \chi_{4}'' + l_{4}(\lambda)\chi_{3}''\right]\left(\sigma^{-} \otimes \sigma^{3}\right), \nonumber   \\
	\fl B(z;\lambda,\mu) = \frac{i}{4 J^{2}}\left[\chi_{1}\chi_{3}' + \chi_{3}\chi_{1}' + \chi_{4}\chi_{2}' +\chi_{2}\chi_{4}' \right] \left(\mathbb{1} \otimes \mathbb{1} \right)  \label{free:B} \\
	\fl + \frac{1}{4 J^{\scriptscriptstyle{3/2}}}\left[-i l_{3}(\mu) \chi_{4}' + l_{4}(\mu)\chi_{3}' \right]\left(\mathbb{1} \otimes \sigma^{-} \right) 
	+ \frac{1}{4 J^{\scriptscriptstyle{3/2}}}\left[-i l_{3}(\mu) \chi_{2}' - l_{4}(\mu)\chi_{1}' \right]\left(\mathbb{1} \otimes \sigma^{+} \right) \nonumber \\
	\fl +\frac{3}{4 J^{\scriptscriptstyle{3/2}}}\left[-i l_{3}(\lambda) \chi_{2}' - l_{4}(\lambda)\chi_{1}' \right]\left(\sigma^{+} \otimes \sigma^{3} \right) 
	+\frac{3}{4 J^{\scriptscriptstyle{3/2}}}\left[-i l_{3}(\lambda) \chi_{4}' + l_{4}(\lambda)\chi_{3}' \right]\left(\sigma^{-} \otimes \sigma^{3} \right), \nonumber \\
	\fl C(z;\lambda,\mu) = \frac{i}{J}\left[ l_{3}(\lambda) l_{3}(\mu) + l_{4}(\lambda) l_{4}(\mu) \right]\left(\sigma^{+} \otimes \sigma^{-} - \sigma^{-} \otimes \sigma^{+} \right) \label{free:C} \\
	\fl + \frac{1}{4 J^{\scriptscriptstyle{3/2}}}\left[-i l_{3}(\mu) \chi_{4} - l_{4}(\mu)\chi_{3} \right]\left(\mathbb{1} \otimes \sigma^{-} \right) 
	+\frac{1}{4 J^{\scriptscriptstyle{3/2}}}\left[i l_{3}(\mu) \chi_{2} + l_{4}(\mu)\chi_{1} \right]\left(\mathbb{1} \otimes \sigma^{+} \right) \nonumber \\
	\fl + \frac{1}{4 J^{\scriptscriptstyle{3/2}}}\left[-i l_{3}(\lambda) \chi_{2} - l_{4}(\mu)\chi_{1} \right]\left(\sigma^{+} \otimes \sigma^{3} \right) 
	+\frac{1}{4 J^{\scriptscriptstyle{3/2}}}\left[-i l_{3}(\lambda) \chi_{4} + l_{4}(\lambda)\chi_{3} \right]\left(\sigma^{-} \otimes \sigma^{3} \right). \nonumber
\end{eqnarray}
Here we have used the canonical structure of the Dirac brackets for the fields $\chi_{i}(z)$, which has an ultralocal form. Nevertheless, the algebra above has a non-ultralocal form. This is exactly the same situation of the AAF model, where despite the ultralocal form of the Dirac brackets structure, one finds a non-ultralocal Lax algebra. We stress, however, that there is a significant simplification when considering the free fermion model instead of the AAF model, since the Dirac brackets structure of the latter has a very complicated, highly non-linear form.

Following the discussion in section \ref{sec:graded_ma}, we have derived the algebra \eref{free:Lax_algebra} using the supertensor product, instead of the usual tensor product. The resulting formulas, \eref{free:A}, \eref{free:B} and \eref{free:C}, however, are written in terms of the usual tensor product. 
 
We now apply our formalism to the infinite line case for sufficiently fast decreasing fields $\chi_{i}(z) \to 0$, $z \to \pm \infty$. In this case, the algebra of transition matrices for equal intervals is given by the appropriate modification of the expression \eref{gma:T_algebra_symm_equal} to the graded case:
\begin{eqnarray} \label{free:T_algebra_symm_equal}
	\left\{ T(x,y;\lambda) \right. \left.\stackrel{\otimes_{s}}{,} T(x,y;\mu) \right\} &=  u(x;\lambda,\mu) \: T(x,y;\lambda) \otimes_{s} T(x,y;\mu) \\
	&- T(x,y;\lambda) \otimes_{s} T(x,y;\mu) \: u(y;\lambda,\mu).\nonumber
\end{eqnarray}
In order to find the local solutions for the $u(x;\lambda,\mu)$-matrix, we follow the prescription given in section \ref{subsec:local}. The calculation is similar to the one given for the $WKIS$ model in section \ref{subsec:WKIS_model}. One must, however, appropriately modify the formulas, taking into account the grading as explained in section \ref{sec:graded_ma}. The equation for the coordinate independent matrix $\tilde{u}(\lambda,\mu)$ takes the form (for additional computational details, see \ref{app:Omega12}):
\begin{equation}
	\left[\tilde{u}(\lambda,\mu),\:\mathcal{L}_{+}(z;\lambda,\mu) \right] = \Omega_{2}(z;\lambda,\mu) - \left[\Omega_{1}(z;\lambda,\mu),\:\mathcal{L}_{+}(z;\lambda,\mu)\right], \label{free:u_tilde}
\end{equation}
where we have used the notation similar to the one used in \eref{local:u_diff_eq}, adapted to the graded case: 
\begin{equation}
	\mathcal{L}_{+}(z;\lambda,\mu):=\left[ \mathcal{L}_1(z;\lambda) \otimes_{s}\mathbb{1} + \mathbb{1} \otimes_{s} \mathcal{L}_1(z;\mu)\right]. 
\end{equation}
We have also split the function $\Omega(z;\lambda,\mu)$ as in \eref{local:Omega_split}: 
\begin{eqnarray*}
\Omega(z;\lambda,\mu) = \partial_{z}\Omega_{1}(z;\lambda,\mu) + \Omega_{2}(z;\lambda,\mu),
\end{eqnarray*}
with the explicit forms of the functions $\Omega_{1}(z;\lambda,\mu)$ and $\Omega_{2}(z;\lambda,\mu)$ given in \ref{app:Omega12}. It is easy to see from the explicit expressions that only with this particular decomposition the matrix $u(x;\lambda,\mu)$ will be a local function of fields. 

It is quite remarkable that, despite the complicated form of the functions $A(z;\lambda,\mu)$, $B(z;\lambda,\mu)$ and $C(z;\lambda,\mu)$ in \eref{free:A}, \eref{free:B} and \eref{free:C}, the solution of the equation \eref{free:u_tilde} exists and has the form:
\begin{eqnarray}\label{free:u_tilde_matrix}
	\tilde{u}(\lambda,\mu) &= \frac{1}{8k} \left[ \sinh(2\lambda + 2 \mu) - 2 \coth(\lambda - \mu) \sinh^2(\lambda+\mu)\right] \mathbb{1}\otimes \sigma_3 \\
	&- \frac{1}{8k} \left[ \sinh(2\lambda + 2 \mu) + 2 \coth(\lambda - \mu) \sinh^2(\lambda+\mu)\right] \sigma_3 \otimes \mathbb{1} \nonumber \\
	&-\frac{\sinh(2 \lambda) \sinh(2 \mu)}{2k \sinh(\lambda-\mu)} (\sigma_+ \otimes \sigma_- + \sigma_- \otimes \sigma_+). \nonumber
\end{eqnarray}
The full $u(z;\lambda,\mu)$-matrix can be obtained from the formula \eref{local:u_local_sol} :
\begin{eqnarray}
	 u(z;\lambda,\mu) = \Omega_{1}(z;\lambda,\mu) + \left[s_{2}(z;\lambda,\mu),\:\mathcal{L}_{+}(z;\lambda,\mu) \right] + \partial_{z}s_{2}(z;\lambda,\mu) \label{free:u_local_sol}\\
	-\frac{1}{2}B(z;\lambda,\mu) + \tilde{u}(\lambda,\mu). \nonumber
\end{eqnarray}
We relegate its explicit expression to \ref{app:Omega12}. The constant part of  the $u(z;\lambda,\mu)$-matrix \eref{local:u_hat_constant}, which defines the algebra of transition matrices for the infinite line limit, with the fields $\chi_{i}(z)$ vanishing at $z \to \pm \infty$, takes the form:
\begin{eqnarray}
	\hat{u}(\lambda,\mu) &= \lim_{z \to \pm \infty} u(z;\lambda,\mu) \label{free:u_hat_constant} \\
	&= \frac{1}{8k} \left[ \sinh(2\lambda + 2 \mu) - 2 \coth(\lambda - \mu) \sinh^2(\lambda+\mu)\right] \mathbb{1}\otimes \sigma_3 \nonumber \\
	&- \frac{1}{8k} \left[ \sinh(2\lambda + 2 \mu) + 2 \coth(\lambda - \mu) \sinh^2(\lambda+\mu)\right] \sigma_3 \otimes \mathbb{1} \nonumber \\
	&+\frac{2 - \cosh(4 \lambda) - \cosh(4 \mu)}{8k \sinh(\lambda-\mu)} (\sigma_+ \otimes \sigma_- + \sigma_- \otimes \sigma_+). \nonumber
\end{eqnarray}
The local $v(z;\lambda,\mu)$-matrix, which will play a role when considering the algebra of  transition matrices for the periodic case, can be easily obtained directly from the equation \eref{gma:v-matrix}, with the appropriate changes made for the graded case. We give its explicit expression in \ref{app:Omega12}.

To summarize, the formalism presented in this paper allows one to obtain a well-defined algebra of transition matrices for both the infinite line limit and the periodic case for the free massive Dirac fermion. It is quite surprising that the Lax pair for such a simple model leads to highly non-ultralocal expressions, which we were able to resolve here using the concept of the $u$- and $v$-matrices, defining the algebra for both cases. Finally, we emphasize that we have obtained this Lax connection from the reduction of the Lax connection for the full $AAF$ model. Note that the latter has a polynomial form with respect to the fields $\chi_{i}(z)$, and, therefore, the matrices $\tilde{u}(\lambda,\mu)$ and  $\hat{u}(\lambda,\mu)$ for the full $AAF$ model should reproduce the corresponding matrices \eref{free:u_tilde_matrix} and \eref{free:u_hat_constant} in the limit $g_{2} \to 0$ and $g_{3} \to 0$.

\subsection{The algebra of transition coefficients}\label{subsec:actang}

We now briefly explain how to derive the algebra of transition coefficients for the free massive fermion model. This is the first essential step towards obtaining the action-angle variables, which we leave for a future publication. As in the previous section, we consider the infinite line case, for which the fields $\chi_{i}(z) \to 0$ when $z \to \pm \infty$. Since the algebra of transition matrices for the infinite line is defined by the $\hat{u}(\lambda,\mu)$-matrix \eref{free:u_hat_constant}, the action-angle variables should be determined from it. Following the monographs \cite{Faddeev:1987ph,Faddeev:1982rn,Novikov:1984id,Korepin:1997bk}, we introduce the reduced monodromy matrix:
\begin{eqnarray}
	T(\lambda) &= \lim_{x \to +\infty} \lim_{y \to -\infty} \left[E_{+}^{-1}(x;\lambda)T(x,y;\lambda)E_{-}(y;\lambda) \right], \label{actang:reduced_mon}
\end{eqnarray}
and denote:
\begin{eqnarray}
T(\lambda)&:= \left( \begin{array}{cccc} 
				t_{1}(\lambda) & t_{2}(\lambda) \\
			t_{3}(\lambda) &  t_{4}(\lambda) 
\end{array}\right),\label{actang:T_t1_t4}
\end{eqnarray}
where $E_{\pm}(x;\lambda)$ are the asymptotical solutions of the differential equation \eref{ov:diff_eq_Tx} in the limits $x \to \pm \infty$:
\begin{eqnarray}
	\left[\partial_{x} - \mathcal{L}_{1}(x;\lambda) \right] E_{\pm}(x;\lambda) = 0, \quad x \to \pm \infty. \label{actang:E_pm}
\end{eqnarray}
Using the explicit expression \eref{free:Lax_pair_L1} for $\mathcal{L}_{1}(x;\lambda)$, we find:
\begin{eqnarray}
	E(x;\lambda):=E_{\pm}(x;\lambda)=e^{\xi_{1}^{(\sigma)}(\lambda)\sigma^{3}x}.
\end{eqnarray}
One then obtains:
\begin{eqnarray}
	\fl \left\{ T(\lambda) \stackrel{\otimes_{s}}{,} T(\mu) \right\} = {u}_{+}(\lambda,\mu) \: T(\lambda) \otimes_{s} T(\mu) - T(\lambda) \otimes_{s} T(\mu) \:{u}_{-}(\lambda,\mu), \label{actang:reduced_mon_alg}
\end{eqnarray}
where the matrices ${u}_{\pm}(\lambda,\mu)$ are defined as follows:
\begin{equation}
	\fl {u}_{\pm}(\lambda,\mu):=\lim_{x \to \pm \infty} \left[E(-x;\lambda) \otimes_{s} E(-x;\mu)\cdot u(x;\lambda,\mu)\cdot E(x;\lambda)\otimes_{s} E(x;\mu)  \right]. \label{actang:u_plus_minus}
\end{equation}

Using the results of the previous section, and the formula for the constant part of $u(z;\lambda,\mu)$ \eref{free:u_hat_constant}, we find:
\begin{eqnarray}
	\fl	u_{+}(\lambda,\mu) &= \left( \begin{array}{cccc} 
			-\text{p.v.}\:a(\lambda,\mu) & 0 & 0 & 0 \\
		 0 &  -b(\lambda,\mu) & i \pi c(\lambda) \delta(\lambda -\mu) & 0 \\
			0 &  -i \pi c(\lambda) \delta(\lambda -\mu) & b(\lambda,\mu) & 0 \\
			0 & 0 & 0 & \text{p.v.}\:a(\lambda,\mu)
			\end{array} \right). \label{actang:u_plus}
\end{eqnarray}
Here, the symbol $\text{p.v.}$ stands for the \emph{principal value}, and we have taken the limit ${x \to \pm \infty}$ in \eref{actang:u_plus_minus} in the sense of the generalized functions, making use of the expressions:
\begin{eqnarray}
 \lim_{x \to +\infty} \text{p.v.}\: \frac{e^{\pm i\alpha(x;\lambda,\mu)}}{\sinh(\lambda - \mu)} &= \mp i \pi \delta(\lambda -\mu), \\
 \lim_{x \to -\infty} \text{p.v.}\: \frac{e^{\pm i\alpha(x;\lambda,\mu)}}{\sinh(\lambda - \mu)} &= \pm i \pi \delta(\lambda -\mu).
\end{eqnarray}
We have also denoted $\alpha(x;\lambda,\mu):=2x\left[\xi_{1}^{(\sigma)}(\lambda) - \xi_{1}^{(\sigma)}(\mu)\right]$. The functions $a(\lambda,\mu)$, $b(\lambda,\mu)$ and $c(\lambda)$ can be easily found from \eref{free:u_hat_constant}:
\begin{eqnarray}
	a(\lambda,\mu) &:= \frac{\coth({\lambda - \mu})\sinh^{2}({\lambda + \mu})}
	{2k}\label{actang:a},\\
	b(\lambda,\mu) &:=\frac{\sinh(2\left(\lambda +\mu)\right)}{4k} \label{actang:b},\\
	c(\lambda) &:=-\frac{1}{2k}\sinh^{2}(2\lambda). \label{actang:c}
\end{eqnarray}
Similarly, we find:
\begin{eqnarray}
	\fl	u_{-}(\lambda,\mu) &= \left( \begin{array}{cccc} 
			-\text{p.v.}\:a(\lambda,\mu) & 0 & 0 & 0 \\
		 0 &  -b(\lambda,\mu) & -i \pi c(\lambda) \delta(\lambda -\mu) & 0 \\
			0 &  i \pi c(\lambda) \delta(\lambda -\mu) & b(\lambda,\mu) & 0 \\
			0 & 0 & 0 & \text{p.v.}\:a(\lambda,\mu)
			\end{array} \right). \label{actang:u_minus}
\end{eqnarray}

It is easy to see from these formulas that the conserved charges, encoded in $\str\left[T(\lambda)\right]=t_{1}(\lambda)-t_{4}(\lambda)$, are in involution:
\begin{equation}
	\left\{ \str\left[T(\lambda)\right], \str\left[T(\mu)\right] \right\} =0. \label{actang:strT_supertrace_zero}
\end{equation}
The conservation of  $\str\left[T(\lambda)\right]$ follows from the time evolution equation for the transition matrix \eref{ov:T_time_evol} and from the appropriate boundary conditions. This can be proved similarly to the bosonic case (see \cite{Faddeev:1987ph,Faddeev:1982rn,Novikov:1984id,Korepin:1997bk} for details).
One can also find the time evolution of the variables $t_{i}(\lambda)$; $i=1,\ldots,4$ from the differential equation for the transition matrix \eref{ov:T_time_evol}:
\begin{eqnarray}
	t_{1}(t;\lambda) &= t_{1}(0,\lambda),\label{actang:t1_time_evol} \\
	t_{2}(t;\lambda) &= t_{2}(0,\lambda)e^{-il_{1}(\lambda)t},\label{actang:t2_time_evol} \\
	t_{3}(t;\lambda) &= t_{3}(0,\lambda)e^{il_{1}(\lambda)t},\label{actang:t3_time_evol} \\
	t_{4}(t;\lambda) &= t_{4}(0,\lambda),\label{actang:t4_time_evol}
\end{eqnarray}
which show that $\str\left[T(\lambda)\right]$ is indeed conserved, moreover, the $t_{1}(t)$ and $t_{4}(t)$ elements are conserved \emph{separately}.

Let us also note, that the components $t_{i}(\lambda)$; $i=1,\ldots,4$ are not independent. Indeed, the Liouville theorem \cite{Berezin:1987ue}, applied to equation \eref{ov:diff_eq_Tx} for the graded case, states that:
\begin{eqnarray}
	\Ber\left[T(x,y;\lambda) \right] = e^{\int_{y}^{x}\str\left[ L(u,y;\lambda) \right]du}, \label{actang:Liouville_th}
\end{eqnarray}
where $\Ber\left[T(x,y;\lambda) \right]$ stands for the \emph{Berezinian} of the matrix $T(x,y;\lambda)$, which for a generic $2 \times 2$ matrix of the type \eref{actang:T_t1_t4} is defined as: $\Ber[T]=(t_1 - t_{2}t_{4}^{-1}t_{3})t_{4}^{-1}$, and satisfies the relation:
\begin{equation}
	\Ber\left[T(x,y;\lambda) \right]=e^{\str\left[T(x,y;\lambda) \right]}. \label{actang:Ber_Str}
\end{equation}
For our case, the formula \eref{actang:Liouville_th} above leads to the following relation between the elements of the monodromy matrix:
\begin{eqnarray}
		\Ber\left[T(x,y;\mu) \right] = e^{\frac{iJl_{2}(\mu)}{k}(x-y)}. \label{actang:Liouville_th_free}
\end{eqnarray}
The Berezinian replaces the concept of the determinant for supermatrices, and the formulas \eref{actang:Liouville_th} and \eref{actang:Ber_Str} are the graded versions of the analogous expressions for the bosonic models. From \eref{actang:Ber_Str}, one obtains the Berezinian for the reduced monodromy matrix \eref{actang:T_t1_t4}:
\begin{eqnarray}
		\Ber\left[T(\lambda) \right] = 1. \label{actang:Ber_reduced_T}
\end{eqnarray}

Using now the formulas \eref{actang:u_plus} and \eref{actang:u_minus}, and writing the algebra \eref{actang:reduced_mon_alg} in components \eref{actang:T_t1_t4}, one then can find the Poisson brackets between the components of the reduced monodromy matrix. The full list of these relations is presented in \ref{app:actang_poisson}. Finally, from these relations, it is easy to read the following brackets:
\begin{eqnarray}
	\left\{ \rho (\lambda ),\rho (\mu )\right\}  &=0, \\
	\left\{ \rho (\lambda ),t_{2}(\mu )\right\}  &=2i\pi t_{2}(\mu )\delta
	\left( \lambda -\mu \right),  \\
	\left\{ \rho (\lambda ),t_{3}(\mu )\right\}  &=-2i\pi t_{3}(\mu )\delta
	\left( \lambda -\mu \right),
\end{eqnarray}
where we have introduced the quantity:
\begin{equation}
	\rho (\lambda ):= \frac{2k }{\sinh ^{2}(2\lambda)}\ln\left( \frac{t_1(\lambda)}{t_4(\lambda)}\right). \label{actang:rho}
\end{equation}
A more careful and detailed study of the classical inverse scattering will be done in a separate publication.

\subsection{Local charges}\label{sec:loccha}

Let us now address the problem of finding the local integrals of motion for the free fermion model in the infinite line (see \cite{Beisert:2005bm} and \cite{Alday:2005gi} for a general discussion for the strings on the $AdS_{5} \times S^{5}$ background). Clearly, there are only three local integrals of motions in this case, corresponding to the momentum $P$, the charge $Q$ and the Hamiltonian $H$ densities:
\begin{eqnarray}
	\fl \mathcal{P}(z) :=(-i) \left[\chi_{3}(z)\chi_{1}'(z) + \chi_{4}(z)\chi_{2}'(z) \right],\label{loccha:P} \\
	\fl \mathcal{Q}(z) := \chi_{3}(z)\chi_{1}(z) + \chi_{4}(z)\chi_{2}(z) ,\label{loccha:Q} \\
	\fl \mathcal{H}(z) := J + \frac{k}{2J} \left[\chi_{4}(z)\chi _{1}^{\prime}(z)-\chi _{3}(z)\chi_{2}^{\prime }(z)-\chi _{4}^{\prime }(z)\chi_{1}(z)+\chi_{3}^{\prime}(z)\chi_{2}(z) \right] \label{loccha:H} \\ 
	- \left[\chi_{4}(z)\chi_{2}(z) - \chi_{3}(z)\chi_{1}(z)\right] \nonumber.
\end{eqnarray}
 It is, therefore, interesting to see how only these three local charges can be extracted from the monodromy \eref{actang:T_t1_t4} in a manner that no other local charges are produced as a result of the expansion of the monodromy.

To explain how this happens, we first write the differential equation for the transition matrix \eref{ov:diff_eq_Tx} in the equivalent integral representation \cite{Faddeev:1987ph,Faddeev:1982rn}. One then can solve the integral equations for the elements of the reduced monodromy matrix $t_{1}(\mu),\ldots,t_{4}(\mu)$ iteratively, order by order in the fields $\chi_{i}$. The final result of this procedure, up to the second order in the fields, takes the form:
\begin{eqnarray}
 t_{1}(\mu) &=1+\int_{-\infty}^{+\infty} \hat{\xi}^{\scriptscriptstyle{(\sigma)}}_{0}(z) dz \label{loccha:t1_2ndorder}\\
&-\int_{-\infty}^{+\infty}e^{2 \hat{\xi}^{\scriptscriptstyle{(\sigma)}}_1 z}\hat{\Lambda}^{\scriptscriptstyle{(+)}}_{\sigma}(z,\mu)\left(
	\int_{z}^{+\infty}e^{-2 \hat{\xi}^{\scriptscriptstyle{(\sigma)}}_1 u}\hat{\Lambda}^{\scriptscriptstyle{(-)}}_{\sigma}(u,\mu)\:du\right) dz, \nonumber \\
 t_{4}(\mu) &=1+\int_{-\infty}^{+\infty}\hat{\xi}^{\scriptscriptstyle{(\sigma)}}_{0}(z)
dz \label{loccha:t4_2ndorder} \\
&-\int_{-\infty}^{+\infty}e^{-2 \hat{\xi}^{\scriptscriptstyle{(\sigma)}}_1 z}\hat{\Lambda}_{\sigma}^{\scriptscriptstyle{(-)}}(z,\mu)\left(
		\int_{z}^{+\infty}e^{2 \hat{\xi}^{\scriptscriptstyle{(\sigma)}}_1 u} \hat{\Lambda}_{\sigma}^{\scriptscriptstyle{(+)}}(u,\mu)\:du\right) dz,\nonumber
\end{eqnarray}
where $\hat{\xi}^{\scriptscriptstyle{(\sigma)}}_{0}(z)$, $\hat{\xi}^{\scriptscriptstyle{(\sigma)}}_{1}(z)$, $\hat{\Lambda}_{\sigma}^{\scriptscriptstyle{(-)}}(z,\mu)$ and $\hat{\Lambda}_{\sigma}^{\scriptscriptstyle{(+)}}(z,\mu)$ are given in \eref{free:xis0}, \eref{free:xis1}, \eref{free:Lambdasm} and \eref{free:Lambdasp}. One can similarly obtain the analogous expressions for the elements $t_{2}(\mu)$ and $t_{3}(\mu)$.

We have already shown in the previous section in \eref{actang:t1_time_evol} and \eref{actang:t4_time_evol} that the components $t_{1}(\mu)$ and $t_{4}(\mu)$ provide the integrals of motion. This can also be checked explicitly, using the equations of motion \eref{free:eoms12} and \eref{free:eoms34}  to verify that the expressions \eref{loccha:t1_2ndorder} and \eref{loccha:t4_2ndorder} are indeed conserved. Moreover, by using the explicit expressions \eref{free:xis0}, \eref{free:xis1}, \eref{free:Lambdasm} and \eref{free:Lambdasp} for  $\hat{\xi}^{\scriptscriptstyle{(\sigma)}}_{0}(z)$, $\hat{\xi}^{\scriptscriptstyle{(\sigma)}}_{1}(z)$, $\hat{\Lambda}_{\sigma}^{\scriptscriptstyle{(-)}}(z,\mu)$ and $\hat{\Lambda}_{\sigma}^{\scriptscriptstyle{(+)}}(z,\mu)$ one can show that the conserved quantity $t_{1}(\mu)$ has the following representation:
\begin{eqnarray}
	t_{1}(\mu) &= - \frac{i}{2}\cosh(2\mu) \int_{-\infty}^{+\infty} \mathcal{P}(z) dz - \frac{i \sinh(2\mu)}{2k}\int_{-\infty}^{+\infty}\mathcal{H}(z)dz \nonumber \\ &+\frac{i}{2k}\cosh(2\mu)\sinh(2\mu)\int_{-\infty}^{+\infty}\mathcal{Q}(z)dz +\sinh^{2}(2\mu)\tilde{t}_{1}(\mu), \label{loccha:t1_local}
\end{eqnarray}
where we have discarded the obvious numerical constants.
Here the densities $\mathcal{P}(z)$, $\mathcal{Q}(z)$ and $\mathcal{H}(z)$ are given in \eref{loccha:P}-\eref{loccha:H}, and we have denoted:
\begin{eqnarray}
	 \tilde{t}_{1}(\mu) &= -\frac{1}{2k}\int_{-\infty}^{+\infty}\left(\chi_{4}\chi_{1}-\chi_{3}\chi_{2}\right)dz \nonumber \\
	&- \frac{J^{2}}{k^{2}} \int_{-\infty}^{+\infty}e^{2 \hat{\xi}^{\scriptscriptstyle{(\sigma)}}_1 z}\Theta^{\scriptscriptstyle{(+)}}(z,\mu)\left(
		\int_{z}^{+\infty}e^{-2 \hat{\xi}^{\scriptscriptstyle{(\sigma)}}_1 u}\Theta^{\scriptscriptstyle{(-)}}(u,\mu)du\right) dz, \label{loccha:t1_tilde}
\end{eqnarray}
where $\Theta^{\scriptscriptstyle{(\pm)}}_{\sigma}(z,\mu)$ are given by the expressions (cf. the formulas \eref{free:Lambdasm} and \eref{free:Lambdasp}):
\begin{eqnarray}
\Theta^{\scriptscriptstyle{(-)}}_{\sigma} &= \frac{1}{\sqrt{J}} \left[ -l_{3}\chi_2- il_{4}\chi_1 \right] \quad \text{and} \quad	\Theta^{\scriptscriptstyle{(+)}}_{\sigma} = \frac{1}{\sqrt{J}} \left[ -l_{3}\chi_4+ il_{4}\chi_3 \right]. \label{loccha:Lambdas_hat_tilde}
\end{eqnarray}
Similar formulas can be obtained for $t_{4}(\mu)$.

Thus, we see from \eref{loccha:t1_local} that the three local charges can be obtained from the decomposition of $t_{1}(\mu)$ around the point $\mu = 0$. In particular, the zeroth and the first orders in $\sinh(2\mu)$ give the three local integrals of motion: the momentum, the Hamiltonian and the charge. Moreover, since $t_{1}(\mu)$ is conserved, it follows from \eref{loccha:t1_local} that the expression $\tilde{t}_{1}(\mu)$ at the next order, $\sinh^{2}(2\mu)$, in \eref{loccha:t1_local} is also conserved. It is also clear from the explicit expressions \eref{loccha:t1_tilde} and \eref{loccha:Lambdas_hat_tilde} that $\tilde{t}_{1}(\mu)$ has a non-local form. Hence, we find that the decomposition of $t_{1}(\mu)$ around the regular point $\mu =0$ produces both local and non-local charges. Moreover, in this manner one obtains exactly three local integrals of motion, corresponding to the momentum, the charge and the Hamiltonian.

We conclude by giving the explicit expression for $\str\left[T(\mu)\right]=t_{1}(\mu)-t_{4}(\mu)$ at the second order in fields:
\begin{eqnarray}
	 \str\left[T(\mu)\right] &= - \frac{J^2\sinh^{2}(2\mu)}{k^2} \left( \int_{-\infty}^{+\infty}e^{2 \hat{\xi}^{\scriptscriptstyle{(\sigma)}}_1 z}\Theta^{\scriptscriptstyle{(+)}}(z,\mu)dz \right) \cdot \label{loccha:str} \\
	&\cdot \left(	\int_{-\infty}^{+\infty}e^{-2 \hat{\xi}^{\scriptscriptstyle{(\sigma)}}_1 u}\Theta^{\scriptscriptstyle{(-)}}(u,\mu)du\right). \nonumber 
\end{eqnarray}
This formula shows that the $\str\left[T(\mu)\right]$ contains only the non-local charges. It is not at this point clear whether the properties of the monodromy matrix considered in this section are the specifics of the free massive fermion model, or some of these features will be also present in the full $AAF$ model.

\section{Conclusion}\label{sec:conclusion}

In this paper we have made the necessary steps towards the quantization of the $AAF$ model, which represents an interesting example of the fermionization technique. More importantly, the results obtained in this paper open the possibility of quantizing the $AAF$ model along the lines of the formalism developed in \cite{Freidel:1991jx,Freidel:1991jv}. In particular, we have shown that one can in principle construct the lattice version of the $AAF$ model, although the quantum version of it is still an open unsolved problem. The same is true for any model which admits a Lax connection with the algebraic structure \eref{gma:Lax_algebra} investigated in this paper. The resulting algebraic structure \eref{gma:lat1}-\eref{gma:lat3} has a more complicated form in comparison to that of the standard integrable models, and requires a more detailed investigation. A particular, simpler case of this algebra has been already considered in \cite{Freidel:1991jx,Freidel:1991jv}.

The key point is that in the process of trading the bosonic fields for fermionic ones in the $AAF$ model, the algebra for the Lax connection becomes highly non-ultralocal, including terms up to the second order derivatives of the delta function. Except for some exotic models, such as the $WKIS$ model, this is a new feature for models obtained from strings. Thus, on the example of the $AAF$ model we have shown how to deal with such models.  Our first central result is the demonstration that the well-defined algebra for transition matrices has the same functional form as in the case considered by Maillet, with the appropriate shift of the $r$- and $s$-matrices. This led us to introduce the shifted dynamical variables, namely, the $u$- and $v$-matrices, which encode the complete information of the scattering data and their dynamics. We have also given a prescription to find the local solutions, and have demonstrated this method by completely solving the $WKIS$ model. 

To make the formalism applicable to the $AAF$ model, we have also given the appropriate generalization to include the graded case. We stress that the formulas given in the main text are general, and can be used for any suitable graded integrable model. To demonstrate our method for the graded case, we have considered a reduction of the $2 \times 2$ representation of the Lax connection for the $AAF$ model, and showed that this reduced Lax connection corresponds to the free massive Dirac fermion model. We have also obtained its corresponding $u$- and $v$-matrices, as well as  derived the algebra of transition coefficients. The problem of extracting the local conserved charges from the monodromy matrix was also addressed. Indeed, they can be obtained, together with the non-local integrals of motion, from the expansion of the monodromy matrix around a regular point.
 

The next step should be obtaining the local $u$- and $v$-matrices for the full $AAF$ model. This is rather a very lengthy and tedious problem, which, however, should not cause any principal problems. The main steps are the same as in the free fermion case. However, it is important to obtain the local solutions and the $u$- and $v$-matrices, which will encode the scattering data for the full model. Moreover, it will be interesting to see to which degree the constant $u$- and $v$-matrices for the full $AAF$ differ from the free massive fermion ones. The relation between the two should shed some light on a possible gauge equivalence between the two models. 

In addition, the results derived in this paper suggests a possible connection between the two formulations of the $\mathfrak{su}(1|1)$ sector of the superstring on $AdS_5 \times S^5$. Let us remind that the $AAF$ model arises in the \emph{uniform} gauge \cite{Alday:2005jm} and leads to very complicated non-linear equations of motion for the fermions \cite{Melikyan:2012kj}. The model is nevertheless classically integrable, and the Lax pair, obtained in \cite{Alday:2005jm} by reduction of the full Lax pair for the superstring on $AdS_{5} \times S^{5}$ background to the $\mathfrak{su}(1|1)$ sector, has a $4 \times 4$ representation. Surprisingly, it was possible to reduce it in \cite{Melikyan:2012kj} to a $2 \times 2$ representation. This is particularly impressive, considering that much simpler fermionic integrable theories, such as the massive Thirring model, do not have a known $2 \times 2$ representation. On the other hand, it is possible to consider the $\mathfrak{su}(1|1)$ sector of the superstring on $AdS_5 \times S^5$ in the so called \emph{uniform light-cone} gauge, where it is described by the free massive fermion theory \cite{Arutyunov:2005hd}.

It can be shown, after an appropriate rescaling of the fields and under the charge conjugation, that the Lax pair for the $\mathfrak{su}(1|1)$ sector of strings in the uniform light-cone gauge coincides with the Lax pair for the free massive fermion model, obtained in section \ref{subsec:free} from the reduction of the $2 \times 2$ Lax connection of the $AAF$ model, which is derived from the $\mathfrak{su}(1|1)$ sector of strings in the uniform gauge. This suggests the possibility of a gauge equivalence between the two Lax connections. Although this is an interesting direction to pursue, one has to consider the quantum version of the AAF model to establish any relation with the free massive Dirac model. This should be possible, due to the formalism presented in this paper, and the study in \cite{Freidel:1991jx,Freidel:1991jv}, where the quantization of the algebras of the form \eref{gma:T_algebra_symm_periodic} has been investigated. This will be considered in details in a future publication.

Finally, we note that it is not immediately clear how to deal with such higher order non-ultralocal terms using the method of \cite{Faddeev:1985qu}. It would be interesting to employ the latter technique for the models with higher order non-ultralocal terms, and find a connection between the two formalisms. This has been an open problem for a long time, and it is especially interesting to explore the  connection between the two techniques in light of the recent progress (see \cite{Delduc:2012qb,Delduc:2012vq} and the references therein) for strings, using the method of \cite{Faddeev:1985qu}. These problems will be considered in a forthcoming publication. 

\ack{The work of A.M. was partially supported by CAPES. The work of G.W. was supported by the FAPESP grant No. 2011/20242-3.}

\section*{Appendices} \addcontentsline{toc}{section}{Appendices}
\appendix 

\section{Computational details for the generalized Maillet algebra}\label{app:gmadetails}

In this appendix we give the computational details of the derivation of the algebra \eref{gma:T_algebra}. Our starting point is the algebra for the $L$-operator given in \eref{gma:L_algebra_s1_s2_v2}. As discussed in the main text, the form of this algebra is not initially obvious to be convenient, but as we show below, it does yield a simple local form for the algebra of transition matrices \eref{gma:T_algebra}. We show this by applying the formula \eref{ov:T_algebra} to each term of \eref{gma:L_algebra_s1_s2_v2}. For convenience we write here the algebra \eref{gma:L_algebra_s1_s2_v2} in the form:
\begin{eqnarray}
	 \fl \left\{ L_1(z;\lambda) \stackrel{\otimes}{,} L_1(z';\mu) \right\} &= \Gamma^{(0)}(z;\lambda,\mu) \: \delta (z-z') + \Gamma^{(1)}(z,z';\lambda,\mu) \: \partial_z \delta (z-z') \label{gma:L_algebra_s1_s2_v2app} \\
	&+ \Gamma^{(2)}(z,z';\lambda,\mu) \: \partial^2_z \delta (z-z') \nonumber,
\end{eqnarray}
where we have denoted:
\begin{eqnarray}
	\fl \Gamma^{(0)}(z;\lambda,\mu):=\partial_{z}r(z;\lambda,\mu) + \left[ r(z;\lambda,\mu), L_1(z;\lambda) \otimes \mathbb{1} + \mathbb{1} \otimes L_1(z;\mu)\right]+ \Lambda(z;\lambda,\mu)\nonumber\\
	\fl \Gamma^{(1)}(z,z';\lambda,\mu):= -\left[s_1(z;\lambda,\mu) + s_1(z';\lambda,\mu)\right]\nonumber\\
	\fl \Gamma^{(2)}(z,z';\lambda,\mu):=s_2(z;\lambda,\mu) + s_2(z';\lambda,\mu).\nonumber
\end{eqnarray}

First, we compute the contribution of the terms proportional to the delta function:
\begin{eqnarray}
	\fl F_{1} :=\int_{y}^{x}dz \int_{y'}^{x'}dz' T(x,z;\lambda) \otimes T(x',z';\mu) \Gamma^{(0)}(z;\lambda,\mu) T(z,y;\lambda) \otimes T(z',y';\mu)\delta(z-z') \nonumber \\
	=T(x,x_{0};\lambda) \otimes T(x',x_{0};\mu) r(x_{0};\lambda,\mu) T(x_{0},y;\lambda) \otimes T(x_{0},y';\mu)\label{gmadetails:F1} \\
	 -T(x,y_{0};\lambda) \otimes T(x',y_{0};\mu) r(y_{0};\lambda,\mu) T(y_{0},y;\lambda) \otimes T(y_{0},y';\mu)\nonumber \\
	+\int_{y_{0}}^{x_{0}}dz\:T(x,z;\lambda) \otimes T(x',z;\mu)\Lambda(z;\lambda,\mu) T(z,y;\lambda) \otimes T(z,y';\mu). \nonumber  
\end{eqnarray}
Here we have also used the notation: 
\begin{equation}
	\mathcal{L}_{\pm}(z;\lambda,\mu):=\mathbb{1} \otimes L_1(z;\mu) \pm L_1(z;\lambda)\otimes\mathbb{1}, \nonumber
\end{equation}
and denoted $x_{0}:=\min(x,x')$ and $y_{0}:=\max(y,y')$.
Next, we find the corresponding expressions for the terms proportional to the first derivative of the delta function in \eref{gma:L_algebra_s1_s2_v2}:
\begin{eqnarray}
	  \fl F_{2} :=\int_{y}^{x}dz\int_{y'}^{x'}dz'\:T(x,z;\lambda) \otimes T(x',z';\mu)  \Gamma^{(1)}(z,z';\lambda,\mu) \label{gmadetails:F2} \\
	\phantom{-}\qquad \qquad \qquad \cdot T(z,y;\lambda) \otimes T(z',y';\mu)\partial_{z}\delta(z-z')\nonumber\\ 
	\fl = T(x,x_{0};\lambda) \otimes T(x',x_{0};\mu) s_{1}(x_{0};\lambda,\mu) T(x_{0},y;\lambda) \otimes T(x_{0},y';\mu)\nonumber \\
	\fl -T(x,y_{0};\lambda) \otimes T(x',y_{0};\mu) s_{1}(y_{0};\lambda,\mu) T(y_{0},y;\lambda) \otimes T(y_{0},y';\mu)\nonumber \\
	\fl -\int_{y_{0}}^{x_{0}}dz\:T(x,z;\lambda) \otimes T(x',z;\mu)  \left[s_{1}(z;\lambda,\mu), \mathcal{L}_{-}(z;\lambda,\mu)\right]T(z,y;\lambda) \otimes T(z,y';\mu).\nonumber
\end{eqnarray}
Finally, the contribution of the terms proportional to the second derivative of the delta function in \eref{gma:Lax_algebra_r_s} can be obtained after somewhat lengthy but straightforward calculations:
\begin{eqnarray}
	\fl F_{3} :=\int_{y}^{x}dz\int_{y'}^{x'}dz'\:T(x,z;\lambda) \otimes T(x',z';\mu)  \left(s_{2}(z;\lambda,\mu) + s_{2}(z';\lambda,\mu) \right)  \nonumber \\
	\qquad \qquad \qquad \cdot T(z,y;\lambda) \otimes T(z',y';\mu)\partial_{z}^{2}\delta(z-z')\label{gmadetails:F3}\\
	\fl =T(x,x_{0};\lambda) \otimes T(x',x_{0};\mu)h_{1}(x_{0};\lambda,\mu)T(x_{0},y;\lambda) \otimes T(x_{0},y';\mu)\nonumber \\
	\fl + T(x,y_{0};\lambda) \otimes T(x',y_{0};\mu) h_{2}(y_{0};\lambda,\mu) T(y_{0},y;\lambda) \otimes T(y_{0},y';\mu)\nonumber \\
	\fl -\int_{y_{0}}^{x_{0}}dz\:T(x,z;\lambda) \otimes T(x',z;\mu)  \cdot h(z;\lambda,\mu) \cdot T(z,y;\lambda) \otimes T(z,y';\mu),\nonumber
\end{eqnarray}
where we have introduced:
\begin{eqnarray}
	\fl h(z;\lambda,\mu) := \left[\partial_{z}s_{2}(z;\lambda,\mu), \mathcal{L}_{+}(z;\lambda,\mu)\right] + \left[ \left[ s_{2}(z;\lambda,\mu) ,\:L_{1}(z;\lambda) \otimes \mathbb{1}  \right],\:\mathbb{1} \otimes L_{1}(z;\mu)\right]\nonumber\\
	+ \left[ \left[ s_{2}(z;\lambda,\mu),\:\mathbb{1} \otimes L_{1}(z;\mu)\right],\:L_{1}(z;\lambda) \otimes \mathbb{1} \right],\label{gmadetails:h_functions}\\
\fl	h_{1}(x_{0};\lambda,\mu):=\partial_{x_{0}}s_{2}(x_{0};\lambda,\mu) + 2\theta(x-x') \left[s_{2}(x_{0};\lambda,\mu), L_{1}(x_{0};\lambda) \otimes \mathbb{1} \right] \nonumber \\
	+ 2\theta(x'-x) \left[s_{2}(x_{0};\lambda,\mu), \mathbb{1} \otimes L_{1}(x_{0};\mu) \right], \nonumber\\
\fl	h_{2}(y_{0};\lambda,\mu):=-\partial_{y_{0}}s_{2}(y_{0};\lambda,\mu) - 2\theta(y-y') \left[s_{2}(x_{0};\lambda,\mu) \mathbb{1} \otimes L_{1}(y_{0};\mu) \right] \nonumber \\
	 - 2\theta(y'-y) \left[s_{2}(y_{0};\lambda,\mu), L_{1}(y_{0};\lambda) \otimes \mathbb{1} \right].\nonumber
\end{eqnarray}

One can see from the formulas \eref{gmadetails:F1}, \eref{gmadetails:F2} and \eref{gmadetails:F3} that the non-local integral terms indeed cancel, if one chooses:
\begin{eqnarray}
  \Lambda(z;\lambda,\mu)=\left[s_{1}(z;\lambda,\mu), \mathcal{L}_{-}(z;\lambda,\mu)\right] +h(z;\lambda,\mu). \label{gmadetails:Lambda_choice_app2}
 \end{eqnarray}
 This function indeed satisfies the requirement discussed in section \ref{sec:gma}, namely, the function $\Lambda(z;\lambda,\mu)$ goes to zero, as the matrices $s_{1}(z;\lambda,\mu)$ and $s_{2}(z;\lambda,\mu)$ go to zero. Collecting the rest of the terms in \eref{gmadetails:F1}, \eref{gmadetails:F2} and \eref{gmadetails:F3} one verifies the local algebra of transition matrices \eref{gma:T_algebra}. For the graded case, all the steps above can be repeated without any changes, replacing everywhere the usual tensor product by the supertensor product.

\section{$2 \times 2$ Lax connection for the $AAF$ model}\label{app:Laxpair}

We use the notations in \cite{Alday:2005jm} for the two-dimensional Dirac matrices in the $AAF$ Lagrangian \eref{aafmodel:Lagrangian}:
\begin{equation}
	\label{dirac_matrices} \rho^0 = \left( 
	\begin{array}{cc}
		-1 & 0 \\
		0 & 1 
	\end{array}
	\right), \quad \rho^1 = \left( 
	\begin{array}{cc}
		0 & i \\
		i & 0 
	\end{array}
	\right)\quad \text{and} \quad \rho^5 = \rho^0 \rho^1.
\end{equation}
The $2 \times 2$ Lax connection for the $AAF$ model is, however, most conveniently written in terms of the fields:
\begin{eqnarray}
	\chi_{1}:=\psi_{1}, \quad \chi_{2}:=\psi_{2}, \quad \chi_{3}:=\psi^{*}_{3} \quad \text{and} \quad \chi_{4}:=\psi^{*}_{2}, \label{notations:chis}
\end{eqnarray}
and the usual Pauli matrices as \cite{Melikyan:2012kj}:
\begin{eqnarray}
		\fl L_{0}(x;\lambda) = \xi_{0}^{\scriptscriptstyle{(\tau)}}(x;\lambda) \mathbb{1} + \xi_{1}^{\scriptscriptstyle{(\tau)}}(x;\lambda) \sigma^{3} + \Lambda^{\scriptscriptstyle{(-)}}_{\tau}(x;\lambda) \sigma^+ + \Lambda^{\scriptscriptstyle{(+)}}_{\tau}(x;\lambda) \sigma^- \label{Laxpair:Lax_pair_L0}, \\
	\fl L_{1}(x;\lambda) = \xi_{0}^{\scriptscriptstyle{(\sigma)}}(x;\lambda) \mathbb{1} + \xi_{1}^{\scriptscriptstyle{(\sigma)}}(x;\lambda) \sigma^{3} + \Lambda^{\scriptscriptstyle{(-)}}_{\sigma}(x;\lambda) \sigma^+ + \Lambda^{\scriptscriptstyle{(+)}}_{\sigma}(x;\lambda) \sigma^-, \label{Laxpair:Lax_pair_L1}
\end{eqnarray}
where the explicit expression of the functions $\xi_{j}^{\scriptscriptstyle{(\sigma,\tau)}}(x;\lambda)$, $j=0,1$ and $\Lambda_{\sigma,\tau}^{\scriptscriptstyle{(\pm)}}(x;\lambda)$ in terms of the components $\chi_{i}$ in \eref{notations:chis} are:\footnote{A more compact way to write these functions was given in \cite{Melikyan:2012kj}. It is convenient, however, to give here the explicit expanded form of the functions.}
\begin{eqnarray}
	\fl \xi^{\scriptscriptstyle{(\sigma)}}_0 = \frac{1}{4J} \left[ - \chi_3\chi_1' + \chi_4\chi_2' - \chi_1\chi_3' + \chi_2\chi_4' \right]\\
	- \frac{1}{4J^2}\left[\chi_2\chi_3\chi_4\chi_1' - \chi_1\chi_3\chi_4\chi_2' - \chi_1\chi_2\chi_4\chi_3' + \chi_1\chi_2\chi_3\chi_4' \right], \nonumber  \\
	\fl \xi^{\scriptscriptstyle{(\sigma)}}_1 = \frac{l_1}{8J} \left[ \chi_3\chi_1' +\chi_4\chi_2' +\chi_1\chi_3' +\chi_2\chi_4' \right] \\ 
	+ \frac{il_2}{4 k} \left[ 2J +\frac{k}{2J} \left( \chi_4\chi_1' - \chi_3\chi_2' + \chi_1\chi_4' - \chi_2\chi_3' \right) +  \left(-\chi_1\chi_3 + \chi_2\chi_4 \right) \right], \nonumber  \\
	\fl \Lambda^{\scriptscriptstyle{(-)}}_{\sigma} = \frac{l_3}{\sqrt{J}} \left[ -\chi_2'  +\frac{1}{4J} \left(2\chi_2\chi_3\chi_1'+ \chi_2\chi_4\chi_2' - \chi_1\chi_3\chi_2' \right) -\frac{1}{16J^2} \chi_1\chi_2\chi_3\chi_4\chi_2' \right]  \\
	+ \frac{i l_4}{\sqrt{J}}  \left[ -\chi_1' + \frac{1}{4J} \left(-2 \chi_1\chi_2\chi_3'+ \chi_2\chi_4\chi_1' - \chi_1\chi_3\chi_1' \right) -\frac{1}{16J^2} \chi_1\chi_2\chi_3\chi_4\chi_1' \right], \nonumber  \\
	\fl \Lambda^{\scriptscriptstyle{(+)}}_{\sigma} = \frac{l_3}{\sqrt{J}} \left[ -\chi_4'  +\frac{1}{4J} \left( 2\chi_1\chi_4\chi_3'+\chi_2\chi_4\chi_4' - \chi_1\chi_3\chi_4' \right) -\frac{1}{16J^2} \chi_1\chi_2\chi_3\chi_4\chi_4' \right]  \\
	+ \frac{i l_4}{\sqrt{J}}  \left[ \chi_3'  +\frac{1}{4J} \left(2\chi_2\chi_3\chi_4'+ \chi_1\chi_3\chi_3' - \chi_2\chi_4\chi_3'  \right) +\frac{1}{16J^2} \chi_1\chi_2\chi_3\chi_4\chi_3' \right], \nonumber
\end{eqnarray}
and:
\begin{eqnarray}
	\xi^{\scriptscriptstyle{(\tau)}}_0 &= \frac{i}{2J} \left[\chi_{3}\chi_{1}+\chi_{4}\chi_{2} \right]+ \frac{1}{4J}\left[-\chi_{3}\dot{\chi}_{1}+\dot{\chi}_{3}\chi_{1} +\chi_{4}\dot{\chi}_{2}  -\dot{\chi}_{4}\chi_{2} \right] \\
	&+ \frac{1}{4J^{2}} \left[-\chi_{2}\chi_{3}\chi_{4}\dot{\chi}_{1} + \chi_{1}\chi_{3}\chi_{4}\dot{\chi}_{2} + \chi_{1}\chi_{2}\chi_{4}\dot{\chi}_{3} - \chi_{1}\chi_{2}\chi_{3}\dot{\chi}_{4} \right], \nonumber \\
	\xi^{\scriptscriptstyle{(\tau)}}_1 &= l_{1}\gamma_{\tau} + \frac{l_{2}k}{8J^{2}} \left[ \chi_{3}{\chi}_{1}' + \chi_{4}{\chi}_{2}' - {\chi}_{3}'{\chi}_{1} - {\chi}_{4}'{\chi}_{2}  \right],\\
	\Lambda^{\scriptscriptstyle{(-)}}_{\tau} &= \frac{2\alpha_{0}\gamma_{\tau}}{\sqrt{J}}\left[l_{3}\chi_{2} - il_{4}\chi_{1} \right] - \frac{1}{\sqrt{J}}\partial_{0} \left[\alpha_{0}\left(l_{3}\chi_{2} + il_{4}\chi_{1} \right) \right], \\
	\Lambda^{\scriptscriptstyle{(+)}}_{\tau} &= -\frac{2\alpha_{0}\gamma_{\tau}}{\sqrt{J}}\left[l_{3}\chi_{4} + il_{4}\chi_{3} \right] - \frac{1}{\sqrt{J}}\partial_{0} \left[\alpha_{0}\left(l_{3}\chi_{4} - il_{4}\chi_{3} \right) \right],
\end{eqnarray}
where:
\begin{eqnarray}
	\fl \gamma_{\tau}:=\frac{1}{8J}\left[ -4iJ + 2i\left(-\chi_{3}\chi_{1} + \chi_{4}\chi_{2} \right) + \chi_{3}\dot{\chi}_{1} + \chi_{4}\dot{\chi}_{2} - \dot{\chi}_{3}{\chi}_{1} - \dot{\chi}_{4}{\chi}_{2}  \right], \label{Laxpair:gamma_tau} \\
	\fl \alpha_{0}:=1+\frac{1}{4 J}\left(-\chi_{3}\chi_{1} +\chi_{4}\chi_{2} \right). \label{Laxpair:alpha0}
\end{eqnarray}
We have also dropped here the dependence on $x$ and on the spectral parameter $\lambda$ to avoid cluttering. The functions $l_{i}$ encoding the dependence on the spectral have the form \cite{Alday:2005jm,Alday:2005gi,Arutyunov:2009ga}:
\begin{eqnarray}
	\fl l_{0} = 1,\; l_{1} = \frac{1+\mu^{2}}{1-\mu^2},\; l_{2} = s_{1}\frac{2\mu}{1-\mu^2}, \;  l_{3} = s_{2}\frac{1}{\sqrt{1-\mu^2}} \; \text{and} \; l_{4} = s_{3}\frac{\mu}{\sqrt{1-\mu^2}}, \label{notations:l_functions}
\end{eqnarray}
where:
\begin{eqnarray}
    s_{1}+s_{2}s_{3} = 0 \quad \text{and} \quad (s_{2})^{2} = (s_{3})^{2} = 1. 
\end{eqnarray}
We also use the alternative parametrization, given by the functions:
\begin{eqnarray}
	\fl l_{0} = 1,\; l_{1} = \cosh{(2\beta)},\; l_{2} = -\sinh{(2\beta)},\; l_{3}=\cosh{(\beta)} \; \text{and} \; l_{4}=\sinh{(\beta)}. \label{notations:alternative_l_functions}
\end{eqnarray}

\section{Computational details for the free massive Dirac fermion model}\label{app:Omega12}

In this appendix, we collect some additional formulas relevant to the computations regarding the free massive Dirac fermion model. First, we give the explicit expressions of the functions $\Omega_{1}(z;\lambda,\mu)$ and $\Omega_{2}(z;\lambda,\mu)$ appearing in the decomposition: 
\begin{eqnarray*}
\Omega(z;\lambda,\mu) = \partial_{z}\Omega_{1}(z;\lambda,\mu) + \Omega_{2}(z;\lambda,\mu).
\end{eqnarray*}
Using the formulas \eref{free:Lambdasm} and \eref{free:Lambdasp}, and introducing the notation:
\begin{eqnarray}
	\hat{\tilde{\Lambda}}^{\scriptscriptstyle{(-)}}_{\sigma} &= \frac{1}{\sqrt{J}} \left[ -l_{3}\chi_2'+ il_{4}\chi_1' \right] \quad \text{and} \quad \hat{\tilde{\Lambda}}^{\scriptscriptstyle{(+)}}_{\sigma} = \frac{1}{\sqrt{J}} \left[ l_{3}\chi_4'+ il_{4}\chi_3' \right],
\end{eqnarray}
one obtains:
\begin{eqnarray}
	\Omega_{1}(z;\lambda,\mu) &=  \frac{i}{8J^{2}}\left[\chi_{1}\chi_{3}' + \chi_{3}\chi_{1}' + \chi_{4}\chi_{2}' +\chi_{2}\chi_{4}'\left(\mathbb{1} \otimes \mathbb{1} \right) \right]\label{Omega12:Omega1} \\
	&+\frac{i}{2J}\hat{\tilde{\Lambda}}^{\scriptscriptstyle{(-)}}_{\sigma}(\lambda)\left(\sigma^{+} \otimes \sigma^{3} \right) -\frac{i}{2J}\hat{\tilde{\Lambda}}^{\scriptscriptstyle{(+)}}_{\sigma}(\lambda)\left(\sigma^{-} \otimes \sigma^{3} \right), \nonumber \\
	\Omega_{2}(z;\lambda,\mu) &= \frac{l_{3}(\lambda)l_{4}(\lambda)}{J^{2}}\left[\chi_{1}'\chi_{4}' + \chi_{2}'\chi_{3}' \right]\left(\mathbb{1} \otimes \mathbb{1} \right) \label{Omega12:Omega2}\\
	&- \frac{4ik}{J} b \:\hat{\xi}_{1}^{\sigma}(\lambda)\hat{\xi}_{1}^{\sigma}(\mu) \left(\sigma^{-} \otimes \sigma^{+} - \sigma^{+} \otimes \sigma^{-} \right) \nonumber\\
	&-\frac{2i}{J}b \: \hat{\xi}_{1}^{\sigma}(\lambda)\hat{\Lambda}^{\scriptscriptstyle{(-)}}_{\sigma}(\mu) \left(\sigma^{+} \otimes \sigma^{3}\right) + \frac{2i}{J} b\: \hat{\xi}_{1}^{\sigma}(\lambda)\hat{\Lambda}^{\scriptscriptstyle{(+)}}_{\sigma}(\mu)\left(\sigma^{-} \otimes \sigma^{3}\right) \nonumber\\
	&- \frac{i}{J} \hat{\xi}_{1}^{\sigma}(\mu)\left[\hat{\tilde{\Lambda}}^{\scriptscriptstyle{(+)}}_{\sigma}(\mu) + 2b \hat{\Lambda}^{\scriptscriptstyle{(+)}}_{\sigma}(\lambda)\right] \left(\mathbb{1} \otimes \sigma^{-}\right) \nonumber \\
	&+ \frac{i}{J} \hat{\xi}_{1}^{\sigma}(\mu)\left[-\hat{\tilde{\Lambda}}^{\scriptscriptstyle{(-)}}_{\sigma}(\mu) + 2b \hat{\Lambda}^{\scriptscriptstyle{(-)}}_{\sigma}(\lambda)\right] \left(\mathbb{1} \otimes \sigma^{+}\right), \nonumber
\end{eqnarray}
where we have denoted $b:=l_{3}(\lambda)l_{3}(\mu) + l_{4}(\lambda)l_{4}(\mu)$.

The equation for the constant part $\tilde{u}(\lambda,\mu)$ of the $u$-matrix has the form \eref{free:u_tilde}:
\begin{equation}
	\left[\tilde{u}(\lambda,\mu),\:\mathcal{L}_{+}(z;\lambda,\mu) \right] = \Gamma(z;\lambda,\mu), \label{Omega12:u_tilde}
\end{equation}
where we have denoted $\Gamma(z;\lambda,\mu):=\Omega_{2}(z;\lambda,\mu) - \left[\Omega_{1}(z;\lambda,\mu),\:\mathcal{L}_{+}(z;\lambda,\mu)\right]$, which in terms of the fields $\chi_{i}(z)$ is given by the following expression:
\begin{eqnarray}
	\Gamma(z;\lambda,\mu) &=- \frac{4 i k}{J} b \: \hat{\xi}_{1}^{\sigma}(\lambda) \hat{\xi}_{1}^{\sigma}(\mu) \left(\sigma^{-} \otimes \sigma^{+} - \sigma^{+} \otimes \sigma^{-} \right)\label{Omega12:Gamma} \\
	&+ \frac{i}{J} \hat{\xi}_{1}^{\sigma}(\lambda)\left[\hat{\tilde{\Lambda}}^{\scriptscriptstyle{(-)}}_{\sigma}(\lambda) - 2b \hat{\Lambda}^{\scriptscriptstyle{(-)}}_{\sigma}(\mu) \right] \left(\sigma^{+} \otimes \sigma^{3} \right)  \nonumber \\
	&- \frac{i}{J} \hat{\xi}_{1}^{\sigma}(\mu)\left[\hat{\tilde{\Lambda}}^{\scriptscriptstyle{(-)}}_{\sigma}(\mu) - 2b \hat{\Lambda}^{\scriptscriptstyle{(-)}}_{\sigma}(\lambda) \right] \left(\mathbb{1} \otimes \sigma^{+} \right) \nonumber \\
	&+ \frac{i}{J} \hat{\xi}_{1}^{\sigma}(\lambda)\left[\hat{\tilde{\Lambda}}^{\scriptscriptstyle{(+)}}_{\sigma}(\lambda) + 2b \hat{\Lambda}^{\scriptscriptstyle{(+)}}_{\sigma}(\mu) \right] \left(\sigma^{-} \otimes \sigma^{3} \right) \nonumber \\
	&-\frac{i}{J}\hat{\xi}_{1}^{\sigma}(\mu)\left[\hat{\tilde{\Lambda}}^{\scriptscriptstyle{(+)}}_{\sigma}(\mu) + 2b \hat{\Lambda}^{\scriptscriptstyle{(+)}}_{\sigma}(\lambda) \right]  \left(\mathbb{1} \otimes \sigma^{-} \right). \nonumber
\end{eqnarray}
The matrix $\mathcal{L}_{+}(z;\lambda,\mu):=\left[ L_1(z;\lambda) \otimes_{s} \mathbb{1} + \mathbb{1} \otimes_{s} L_1(z;\mu)\right]$ has the form:
\begin{eqnarray}
	\fl \mathcal{L}_{+}(z;\lambda,\mu)=\left[ \hat{\xi}_{1}^{\sigma}(\lambda) + \hat{\xi}_{1}^{\sigma}(\mu) \right]\left(\mathbb{1} \otimes \mathbb{1} \right) + \hat{\xi}_{1}^{\sigma}(\lambda)\left(\sigma^{3} \otimes \mathbb{1} \right) + \hat{\xi}_{1}^{\sigma}(\mu)\left(\mathbb{1} \otimes \sigma^{3} \right)\\
	 +\hat{\Lambda}^{\scriptscriptstyle{(-)}}_{\sigma}(\lambda)\left(\sigma^{+} \otimes \sigma^{3} \right)+\hat{\Lambda}^{\scriptscriptstyle{(-)}}_{\sigma}(\mu)\left(\mathbb{1} \otimes \sigma^{+} \right)
	+\hat{\Lambda}^{\scriptscriptstyle{(+)}}_{\sigma}(\lambda)\left(\sigma^{-} \otimes \sigma^{3} \right)+\hat{\Lambda}^{\scriptscriptstyle{(+)}}_{\sigma}(\mu)\left(\mathbb{1} \otimes \sigma^{-} \right). \nonumber
\end{eqnarray}

Finally, we give the explicit expressions for the $u(z;\lambda,\mu)$ and $v(z;\lambda,\mu)$ matrices for the free massive fermion. The former can be easily computed from \eref{free:u_local_sol},
\begin{eqnarray} \label{Omega12:u-matrix}
	\fl u(z;\lambda,\mu) = \tilde{u}(\lambda,\mu) + \frac{1}{8J^2} \left[ \left( \cosh \lambda \sinh \lambda - \cosh \mu \cosh \mu \right) \partial_z \left( \chi_1 \chi_4 + \chi_2 \chi_3 \right) \right. \\ 
	- \left. i \left( \sinh^2 \lambda - \sinh^2 \mu \right) \left( \chi_1 \chi_3' + \chi_3 \chi_1' \right) + i \left( \cosh^2 \lambda - \cosh^2 \mu \right) \left( \chi_2 \chi_4' + \chi_4 \chi_2' \right) \right] \left( \mathbb{1} \otimes \mathbb{1} \right)\nonumber  \\
	- \frac{1}{k} \cosh^2(\lambda+\mu) \sinh(\lambda-\mu) \left( \sigma^+ \otimes \sigma^- + \sigma^- \otimes \sigma^+ \right) \nonumber \\
	+\left\{ \frac{\sinh (2\mu)}{8k\sqrt{J}} \left[ -\cosh \mu \: \chi_2 + i \sinh \mu \: \chi_1 \right]  \right. \nonumber \\
	\qquad + \left. \frac{1}{4J^\frac{3}{2}} \left[ -i \cosh(2 \lambda +\mu) \chi_2' + \sinh(2\lambda + \mu) \chi_1' \right] \right\} \left( \mathbb{1} \otimes \sigma^+ \right) \nonumber   \\
	+\left\{ \frac{\sinh (2\mu)}{8k\sqrt{J}} \left[ \cosh \mu \: \chi_4 + i \sinh \mu \: \chi_3 \right]  \right. \nonumber \\
	\qquad + \left. \frac{1}{4J^\frac{3}{2}} \left[ -i \cosh(2 \lambda +\mu) \chi_4' - \sinh(2\lambda + \mu) \chi_3' \right] \right\} \left( \mathbb{1} \otimes \sigma^- \right) \nonumber  \\
	+\left\{ \frac{\sinh (2\lambda)}{8k\sqrt{J}} \left[ \cosh \lambda \: \chi_2 - i \sinh \lambda \: \chi_1 \right]  \right. \nonumber \\
	\qquad + \left. \frac{1}{4J^\frac{3}{2}} \left[ i \cosh( \lambda + 2\mu) \chi_2' - \sinh(\lambda + 2\mu) \chi_1' \right] \right\} \left( \sigma^+ \otimes \sigma^3 \right) \nonumber  \\
	+\left\{ \frac{\sinh (2\lambda)}{8k\sqrt{J}} \left[ - \cosh \lambda \: \chi_4 - i \sinh \lambda \: \chi_3 \right]  \right. \nonumber \\
	\qquad + \left. \frac{1}{4J^\frac{3}{2}} \left[ i \cosh( \lambda + 2\mu) \chi_4' + \sinh(\lambda + 2\mu) \chi_3' \right] \right\} \left( \sigma^- \otimes \sigma^3 \right), \nonumber
\end{eqnarray}
where the constant $\tilde{u}(\lambda,\mu)$ is given by \eref{free:u_tilde_matrix}. While the latter can be obtained from the graded counterpart of \eref{gma:v-matrix},
\begin{eqnarray} \label{Omega12:v-matrix}
	\fl v(z;\lambda,\mu) =  \frac{1}{8J^2} \left[ \frac{1}{2}\left[ \sinh (2 \lambda) + \sinh (2 \mu) \right] \partial_z \left( \chi_1 \chi_4 + \chi_2 \chi_3 \right) \right. \\ 
	+ \left. i \left( \cosh^2 \lambda + \sinh^2 \mu \right) \left( \chi_4 \chi_2' + \chi_2 \chi_4' \right) - i \left( \cosh^2 \mu + \sinh^2 \lambda \right) \left( \chi_3 \chi_1' + \chi_1 \chi_3' \right) \right] \left( \mathbb{1} \otimes \mathbb{1} \right)\nonumber  \\
	- \frac{1}{2k} \cosh(\lambda+\mu) \left[ \sinh 2\lambda + \sinh 2\mu \right] \left( \sigma^+ \otimes \sigma^- + \sigma^- \otimes \sigma^+ \right) \nonumber  \\
	+\left\{ \frac{\sinh (2\mu)}{8k\sqrt{J}} \left[ \cosh \mu \: \chi_2 - i \sinh \mu \: \chi_1 \right]  \right. \nonumber \\
	\qquad + \left. \frac{1}{4J^\frac{3}{2}} \left[ -i \cosh(2 \lambda +\mu) \chi_2' + \sinh(2\lambda + \mu) \chi_1' \right] \right\} \left( \mathbb{1} \otimes \sigma^+ \right) \nonumber  \\
	+\left\{ \frac{\sinh (2\mu)}{8k\sqrt{J}} \left[ - \cosh \mu \: \chi_4 - i \sinh \mu \: \chi_3 \right]  \right. \nonumber \\
	\qquad + \left. \frac{1}{4J^\frac{3}{2}} \left[ -i \cosh(2 \lambda +\mu) \chi_4' - \sinh(2\lambda + \mu) \chi_3' \right] \right\} \left( \mathbb{1} \otimes \sigma^- \right) \nonumber  \\
	+\left\{ \frac{\sinh (2\lambda)}{8k\sqrt{J}} \left[ \cosh \lambda \: \chi_2 - i \sinh \lambda \: \chi_1 \right]  \right. \nonumber \\
	\qquad + \left. \frac{1}{4J^\frac{3}{2}} \left[ - i \cosh( \lambda + 2\mu) \chi_2' + \sinh(\lambda + 2\mu) \chi_1' \right] \right\} \left( \sigma^+ \otimes \sigma^3 \right) \nonumber  \\
	+\left\{ \frac{\sinh (2\lambda)}{8k\sqrt{J}} \left[ - \cosh \lambda \: \chi_4 - i \sinh \lambda \: \chi_3 \right]  \right. \nonumber \\
	\qquad + \left. \frac{1}{4J^\frac{3}{2}} \left[- i \cosh( \lambda + 2\mu) \chi_4' - \sinh(\lambda + 2\mu) \chi_3' \right] \right\} \left( \sigma^- \otimes \sigma^3 \right). \nonumber
\end{eqnarray}

\section{Poisson brackets for the reduced monodromy elements}\label{app:actang_poisson}

In this appendix we give the list of the Poisson brackets for the elements of the reduced monodromy matrix \eref{actang:T_t1_t4}:
\begin{enumerate}
\item $\left\{ t_{1}(\lambda ),t_{1}(\mu )\right\} =0$,
\item $\left\{ t_{1}(\lambda ),t_{4}(\mu )\right\} =0$,
\item $\left\{ t_{4}(\lambda ),t_{4}(\mu )\right\} =0$,
\item $\left\{ t_{2}(\lambda ),t_{2}(\mu )\right\} =-2 \left(\text{p.v.}\:a(\lambda,\mu)\right)t_{2}(\lambda
)t_{2}(\mu )$,
\item $\left\{ t_{3}(\lambda ),t_{3}(\mu )\right\} =2\left(\text{p.v.}\:a(\lambda,\mu)\right)t_{3}(\lambda
)t_{3}(\mu )$,
\item $\left\{ t_{1}(\lambda ),t_{2}(\mu )\right\} =-\left[ \left(\text{p.v.}\:a(\lambda,\mu)\right)-b(\lambda,\mu)\right]
t_{1}(\lambda )t_{2}(\mu )-i \pi c(\lambda,\mu)t_{2}(\lambda )t_{1}(\mu )\delta \left(\lambda -\mu \right)$,
\item $\left\{ t_{1}(\lambda ),t_{3}(\mu )\right\} =\left[ \left(\text{p.v.}\:a(\lambda,\mu)\right)-b(\lambda,\mu)\right]
t_{1}(\lambda )t_{3}(\mu )+i \pi c(\lambda,\mu)t_{3}(\lambda )t_{1}(\mu )\delta \left(\lambda -\mu \right)$,
\item $\left\{ t_{2}(\lambda ),t_{3}(\mu )\right\} =-2b(\lambda,\mu)t_{2}(\lambda
)t_{3}(\mu )-2i \pi c(\lambda,\mu)t_{1}(\lambda )t_{4}(\mu )\delta \left( \lambda -\mu
\right)$,
\item $\left\{ t_{2}(\lambda ),t_{4}(\mu )\right\} =-\left[ \left(\text{p.v.}\:a(\lambda,\mu)\right)+b(\lambda,\mu)\right]
t_{2}(\lambda )t_{4}(\mu )-i \pi c(\lambda,\mu)t_{4}(\lambda )t_{2}(\mu )\delta \left(
\lambda -\mu \right)$,
\item $\left\{ t_{3}(\lambda ),t_{4}(\mu )\right\} =\left[ \left(\text{p.v.}\:a(\lambda,\mu)\right)+b(\lambda,\mu)\right]
t_{3}(\lambda )t_{4}(\mu )+i \pi c(\lambda,\mu)t_{4}(\lambda )t_{3}(\mu )\delta \left(
\lambda -\mu \right)$.
\end{enumerate}

 \newpage
 
 \section*{References}
\bibliographystyle{utphys} 
\bibliography{AAFMaillet_arxiv_v2}

\providecommand{\href}[2]{#2}\begingroup\raggedright\begin{thebibliography}{10}

\bibitem{Alday:2005jm}
L.~F. Alday, G.~Arutyunov, and S.~Frolov, ``{New integrable system of 2dim
  fermions from strings on $AdS_5 \times S^5$},''
  \href{http://dx.doi.org/10.1088/1126-6708/2006/01/078}{{\em JHEP} {\bf 0601}
  (2006)  078},
\href{http://arxiv.org/abs/hep-th/0508140}{{\tt arXiv:hep-th/0508140}}.

\bibitem{Arutyunov:2005hd}
G.~Arutyunov and S.~Frolov, ``{Uniform light-cone gauge for strings in $AdS_5
  \times S^5$: Solving $su(1|1)$ sector},''
  \href{http://dx.doi.org/10.1088/1126-6708/2006/01/055}{{\em JHEP} {\bf 0601}
  (2006)  055},
\href{http://arxiv.org/abs/hep-th/0510208}{{\tt arXiv:hep-th/0510208}}.

\bibitem{Staudacher:2004tk}
M.~Staudacher, ``{The Factorized S-matrix of CFT/AdS},''
  \href{http://dx.doi.org/10.1088/1126-6708/2005/05/054}{{\em JHEP} {\bf 0505}
  (2005)  054},
\href{http://arxiv.org/abs/hep-th/0412188}{{\tt arXiv:hep-th/0412188}}.

\bibitem{Frolov:2006cc}
S.~Frolov, J.~Plefka, and M.~Zamaklar, ``The ${AdS_{5} \times S^{5}}$
  superstring in light-cone gauge and its bethe equations,'' {\em J. Phys.}
  {\bf A39} (2006)  13037--13082,
\href{http://arxiv.org/abs/hep-th/0603008}{{\tt hep-th/0603008}}.

\bibitem{Arutyunov:2006ak}
G.~Arutyunov, S.~Frolov, J.~Plefka, and M.~Zamaklar, ``{The Off-shell Symmetry
  Algebra of the Light-cone $AdS_{5} \times S^{5}$ Superstring},''
  \href{http://dx.doi.org/10.1088/1751-8113/40/13/018}{{\em J.Phys.} {\bf A40}
  (2007)  3583--3606},
\href{http://arxiv.org/abs/hep-th/0609157}{{\tt arXiv:hep-th/0609157}}.

\bibitem{McLoughlin:2004dh}
T.~McLoughlin and I.~Swanson, ``{N-impurity superstring spectra near the
  pp-wave limit},''
  \href{http://dx.doi.org/10.1016/j.nuclphysb.2004.09.025}{{\em Nucl.Phys.}
  {\bf B702} (2004)  86--108},
\href{http://arxiv.org/abs/hep-th/0407240}{{\tt arXiv:hep-th/0407240
  [hep-th]}}.

\bibitem{Swanson:2005wz}
I.~Swanson, ``{Superstring holography and integrability in $AdS_{5} \times
  S^{5}$},''
\href{http://arxiv.org/abs/hep-th/0505028}{{\tt arXiv:hep-th/0505028
  [hep-th]}}.

\bibitem{Klose:2006dd}
T.~Klose and K.~Zarembo, ``{Bethe ansatz in stringy sigma models},''
  \href{http://dx.doi.org/10.1088/1742-5468/2006/05/P05006}{{\em J.Stat.Mech.}
  {\bf 0605} (2006)  P05006},
\href{http://arxiv.org/abs/hep-th/0603039}{{\tt arXiv:hep-th/0603039
  [hep-th]}}.

\bibitem{Melikyan:2011uf}
A.~Melikyan, A.~Pinzul, V.~Rivelles, and G.~Weber, ``{Quantum integrability of
  the Alday-Arutyunov-Frolov model},''
  \href{http://dx.doi.org/10.1007/JHEP09(2011)092}{{\em JHEP} {\bf 1109} (2011)
   092},
\href{http://arxiv.org/abs/1106.0512}{{\tt arXiv:1106.0512 [hep-th]}}.

\bibitem{Melikyan:2012kj}
A.~Melikyan and G.~Weber, ``{The r-matrix of the Alday-Arutyunov-Frolov
  model},'' \href{http://dx.doi.org/10.1007/JHEP11(2012)165}{{\em JHEP} {\bf
  1211} (2012)  165},
\href{http://arxiv.org/abs/1209.6042}{{\tt arXiv:1209.6042 [hep-th]}}.

\bibitem{Beisert:2010jr}
N.~Beisert {\em et al.}, ``{Review of AdS/CFT Integrability: An Overview},''
  \href{http://dx.doi.org/10.1007/s11005-011-0529-2}{{\em Lett.Math.Phys.} {\bf
  99} (2012)  3--32},
\href{http://arxiv.org/abs/1012.3982}{{\tt arXiv:1012.3982 [hep-th]}}.

\bibitem{Arutyunov:2009ga}
G.~Arutyunov and S.~Frolov, ``{Foundations of the $AdS_5 \times S^5$
  Superstring. Part I},''
  \href{http://dx.doi.org/10.1088/1751-8113/42/25/254003}{{\em J. Phys.} {\bf
  A42} (2009)  254003},
\href{http://arxiv.org/abs/0901.4937}{{\tt arXiv:0901.4937 [hep-th]}}.

\bibitem{Polyakov:1983tt}
A.~M. Polyakov and P.~B. Wiegmann, ``{Theory of nonabelian {G}oldstone bosons
  in two dimensions},''
\href{http://dx.doi.org/10.1016/0370-2693(83)91104-8}{{\em Phys. Lett.} {\bf
  B131} (1983)  121--126}.

\bibitem{Polyakov:1984et}
A.~M. Polyakov and P.~Wiegmann, ``{Goldstone Fields in Two-Dimensions with
  Multivalued Actions},''
\href{http://dx.doi.org/10.1016/0370-2693(84)90206-5}{{\em Phys.Lett.} {\bf
  B141} (1984)  223--228}.

\bibitem{Faddeev:1985qu}
L.~D. Faddeev and N.~Y. Reshetikhin, ``Integrability of the principal chiral
  field model in (1+1) - dimension,''
\href{http://dx.doi.org/10.1016/0003-4916(86)90201-0}{{\em Ann. Phys.} {\bf
  167} (1986)  227}.

\bibitem{deVega:1983gy}
H.~J. de~Vega, H.~Eichenherr, and J.~M. Maillet, ``Classical and quantum
  algebras of nonlocal charges in sigma models,''
\href{http://dx.doi.org/10.1007/BF01215281}{{\em Commun. Math. Phys.} {\bf 92}
  (1984)  507}.

\bibitem{Maillet:1985ec}
J.~M. Maillet, ``{Hamiltonian structures for integrable classical theories from
  graded Kac-Moody algebras},''
\href{http://dx.doi.org/10.1016/0370-2693(86)91289-X}{{\em Phys.Lett.} {\bf
  B167} (1986)  401}.

\bibitem{Kundu:2003cu}
A.~Kundu, ``Unifying approaches in integrable systems: Quantum and statistical,
  ultralocal and nonultralocal,'' in {\em Classical and Quantum Nonlinear
  Integrable Systems}, pp.~147--181.
\newblock IOP Publishing, Bristol, UK, 2003.

\bibitem{Kundu:1996hb}
A.~Kundu, ``{Quantum integrable systems: Construction, solution, algebraic
  aspect},''
\href{http://arxiv.org/abs/hep-th/9612046}{{\tt arXiv:hep-th/9612046
  [hep-th]}}.

\bibitem{Delduc:2012qb}
F.~Delduc, M.~Magro, and B.~Vicedo, ``{Alleviating the non-ultralocality of
  coset sigma models through a generalized Faddeev-Reshetikhin procedure},''
  \href{http://dx.doi.org/10.1007/JHEP08(2012)019}{{\em JHEP} {\bf 1208} (2012)
   019},
\href{http://arxiv.org/abs/1204.0766}{{\tt arXiv:1204.0766 [hep-th]}}.

\bibitem{Delduc:2012vq}
F.~Delduc, M.~Magro, and B.~Vicedo, ``{Alleviating the non-ultralocality of the
  $AdS_5 \times S^5$ superstring},''
  \href{http://dx.doi.org/10.1007/JHEP10(2012)061}{{\em JHEP} {\bf 1210} (2012)
   061},
\href{http://arxiv.org/abs/1206.6050}{{\tt arXiv:1206.6050 [hep-th]}}.

\bibitem{Dorey:2006mx}
N.~Dorey and B.~Vicedo, ``{A Symplectic Structure for String Theory on
  Integrable Backgrounds},''
  \href{http://dx.doi.org/10.1088/1126-6708/2007/03/045}{{\em JHEP} {\bf 0703}
  (2007)  045},
\href{http://arxiv.org/abs/hep-th/0606287}{{\tt arXiv:hep-th/0606287
  [hep-th]}}.

\bibitem{Benichou:2010ts}
R.~Benichou, ``{Fusion of line operators in conformal sigma-models on
  supergroups, and the Hirota equation},''
  \href{http://dx.doi.org/10.1007/JHEP01(2011)066}{{\em JHEP} {\bf 1101} (2011)
   066},
\href{http://arxiv.org/abs/1011.3158}{{\tt arXiv:1011.3158 [hep-th]}}.

\bibitem{Benichou:2011ch}
R.~Benichou, ``{First-principles derivation of the AdS/CFT Y-systems},''
  \href{http://dx.doi.org/10.1007/JHEP10(2011)112}{{\em JHEP} {\bf 1110} (2011)
   112},
\href{http://arxiv.org/abs/1108.4927}{{\tt arXiv:1108.4927 [hep-th]}}.

\bibitem{Benichou:2012hc}
R.~Benichou, ``{The Hirota equation for string theory in $AdS_5 \times S^5$
  from the fusion of line operators},''
  \href{http://dx.doi.org/10.1002/prop.201200024}{{\em Fortsch.Phys.} {\bf 60}
  (2012)  896--900},
\href{http://arxiv.org/abs/1202.0084}{{\tt arXiv:1202.0084 [hep-th]}}.

\bibitem{Maillet:1985fn}
J.~M. Maillet, ``{Kac-Moody algebra and extended Yang-Baxter relations in the
  {O(n)} nonlinear sigma model},''
\href{http://dx.doi.org/10.1016/0370-2693(85)91075-5}{{\em Phys.Lett.} {\bf
  B162} (1985)  137}.

\bibitem{Maillet:1985ek}
J.~M. Maillet, ``New integrable canonical structures in two-dimensional
  models,''
\href{http://dx.doi.org/10.1016/0550-3213(86)90365-2}{{\em Nucl. Phys.} {\bf
  B269} (1986)  54}.

\bibitem{Freidel:1991jx}
L.~Freidel and J.~Maillet, ``{Quadratic algebras and integrable systems},''
\href{http://dx.doi.org/10.1016/0370-2693(91)91566-E}{{\em Phys.Lett.} {\bf
  B262} (1991)  278--284}.

\bibitem{Freidel:1991jv}
L.~Freidel and J.~Maillet, ``{On classical and quantum integrable field
  theories associated to Kac-Moody current algebras},''
\href{http://dx.doi.org/10.1016/0370-2693(91)90479-A}{{\em Phys. Lett.} {\bf
  B263} (1991)  403--410}.

\bibitem{Faddeev:1987ph}
L.~D. Faddeev and L.~A. Takhtajan, {\em Hamiltonian Methods in the Theory of
  Solitons}.
\newblock Springer Series in Soviet Mathematics, 592 p., Springer-Verlag Berlin
  Heidelberg, 1987.

\bibitem{Faddeev:1982rn}
L.~D. Faddeev, ``Integrable models in (1+1)-dimensional quantum field theory,''
  {\em Recent advances in Field Theory and Statistical Mechanics} (1982)  .
  Proc. of Summer School of Theoretical Physics, Les Houches, France, eds. J.B.
  Zuber and R. Stora (North-Holland, Amsterdam, 1984).

\bibitem{Novikov:1984id}
S.~Novikov, S.~V. Manakov, L.~P. Pitaevsky, and V.~E. Zakharov, {\em Theory of
  Solitons : The Inverse Scattering Method, 276 p.}
\newblock Contemporary Soviet Mathematics. New York, USA: Consultants Bureau,
  1984.

\bibitem{Korepin:1997bk}
V.~E. Korepin, N.~M. Bogoliubov, and A.~G. Izergin, {\em Quantum Inverse
  Scattering Method and Correlation Functions}.
\newblock Cambridge Monographs on Mathematical Physics. Cambridge University
  Press, 1997.

\bibitem{wadati:1980ts}
T.~Shimizu and M.~Wadati, ``A new integrable nonlinear evolution equation,''
  \href{http://dx.doi.org/10.1143/PTP.63.808}{{\em Prog. Theor. Phys.} {\bf 63}
  (1980) no.~3, 808--820}.

\bibitem{Wadati:1965ff}
M.~Wadati, K.~Konno, and Y.-H. Ichikawa, ``{A Generalization of Inverse
  Scattering Method},'' \href{http://dx.doi.org/10.1143/JPSJ.46.1965}{{\em
  Journal of the Physical Society of Japan} {\bf 46} (1979) no.~6, 1965--1966}.

\bibitem{calogero:1982de}
F.~Calogero and A.~Degasperis, {\em {Spectral Transform and Solitons One:
  Spectral Transform and Solitons: Tools to Solve and Investigate Nonlinear
  Evolution Equations.}}
\newblock New York: North-Holland, 1982.

\bibitem{Tsyplyaev:1981cz}
S.~Tsyplyaev, ``{Commutation relations of the transition matrix in the
  classical and quantum inverse scattering methods (local case)},''
\href{http://dx.doi.org/10.1007/BF01037981}{{\em Theor. Math. Phys.} {\bf 48}
  (1982)  580--586}.

\bibitem{Gorder:2012rv}
R.~A.~V. Gorder, ``{Exact Stationary Solution Method for the
  Wadati-Konno-Ichikawa-Shimizu (WKIS) Equation},''
  \href{http://dx.doi.org/10.1143/PTP.128.993}{{\em Progress of Theoretical
  Physics} {\bf 128} (2012) no.~5, 993--999}.

\bibitem{Lakshmanan:1985:gs}
M.~Lakshmanan and S.~Ganesan, ``{Geometrical and gauge equivalence of the
  generalized Hirota, Heisenberg and Wkis equations with linear
  inhomogeneities},''
  \href{http://dx.doi.org/10.1016/0378-4371(85)90120-7}{{\em Physica A:
  Statistical Mechanics and its Applications} {\bf 132} (1985) no.~1,
  117--142}.

\bibitem{Lakshmanan:1983sg}
M.~Lakshmanan and S.~Ganesan, ``{Equivalent forms of a generalized Hirota's
  equation with linear inhomogeneities},''
  \href{http://dx.doi.org/10.1143/JPSJ.52.4031}{{\em Journal of the Physical
  Society of Japan} {\bf 52} (1983) no.~12, 4031--4033}.

\bibitem{Alday:2005gi}
L.~F. Alday, G.~Arutyunov, and A.~A. Tseytlin, ``{On integrability of classical
  superstrings in $AdS_5 \times S^5$},''
  \href{http://dx.doi.org/10.1088/1126-6708/2005/07/002}{{\em JHEP} {\bf 0507}
  (2005)  002},
\href{http://arxiv.org/abs/hep-th/0502240}{{\tt arXiv:hep-th/0502240}}.

\bibitem{Berezin:1987ue}
F.~Berezin, {\em {Introduction to Superanalysis}}.
\newblock Mathematical Physics and Applied Mathematics. Springer, 1987.

\bibitem{Kulish:1980ii}
P.~Kulish and E.~Sklyanin, ``{On the solution of the Yang-Baxter equation},''
\href{http://dx.doi.org/10.1007/BF01091463}{{\em J. Sov. Math.} {\bf 19} (1982)
   1596--1620}.

\bibitem{Kulish:1985bj}
P.~Kulish, ``{Integrable graded magnets},''
\href{http://dx.doi.org/10.1007/BF01083770}{{\em J. Sov. Math.} {\bf 35} (1986)
   2648--2662}.

\bibitem{Gohmann:1998sm}
F.~G{\"o}hmann and S.~Murakami, ``Fermionic representations of integrable
  lattice systems,'' \href{http://dx.doi.org/10.1088/0305-4470/31/38/009}{{\em
  Journal of Physics A: Mathematical and General} {\bf 31} (1998) no.~38,
  7729}.

\bibitem{Gohmann:1999av}
F.~Gohmann and V.~Korepin, ``{Solution of the quantum inverse problem},''
  \href{http://dx.doi.org/10.1088/0305-4470/33/6/308}{{\em J.Phys.} {\bf A33}
  (2000)  1199--1220},
\href{http://arxiv.org/abs/hep-th/9910253}{{\tt arXiv:hep-th/9910253
  [hep-th]}}.

\bibitem{Gohman:2002kp}
F.~G{\"o}hmann and V.~Korepin, ``{A quantum version of the inverse scattering
  transformation},'' \href{http://dx.doi.org/10.1134/1.1490094}{{\em Physics of
  Atomic Nuclei} {\bf 65} (2002) no.~6, 968--975}.

\bibitem{Essler:2005bk}
F.~H.~L. Essler, H.~Frahm, F.~G{\"o}hmann, A.~Kl{\"u}mper, and V.~E. Korepin,
  \href{http://dx.doi.org/10.1017/CBO9780511534843}{{\em The One-Dimensional
  {H}ubbard Model}}.
\newblock Cambridge University Press, Cambridge, 2005.

\bibitem{Beisert:2005bm}
N.~Beisert, V.~Kazakov, K.~Sakai, and K.~Zarembo, ``{The Algebraic curve of
  classical superstrings on $AdS_{5} \times S^{5}$},''
  \href{http://dx.doi.org/10.1007/s00220-006-1529-4}{{\em Commun.Math.Phys.}
  {\bf 263} (2006)  659--710},
\href{http://arxiv.org/abs/hep-th/0502226}{{\tt arXiv:hep-th/0502226
  [hep-th]}}.

\end{thebibliography}\endgroup

\end{document}